\documentclass{aa}  

\usepackage{anyfontsize}
\usepackage{tensor} 
\usepackage{xargs}
\usepackage{xspace}
\usepackage{textgreek}

\newcommand{\mum}{\text{\textmu m}\xspace}
\newcommand{\Lyalpha}{\text{Ly\textalpha}\xspace}
\newcommand{\Halpha}{\text{H\textalpha}\xspace}
\newcommand{\Hbeta}{\text{H\textbeta}\xspace}
\newcommand{\Hgamma}{\text{H\textgamma}\xspace}
\newcommand{\Hdelta}{\text{H\textdelta}\xspace}

\newcommand{\Hepsilon}{\text{H\textepsilon}\xspace}
\newcommand{\Hzeta}{\text{H\textzeta}\xspace}

\newcommandx{\permittedEL}[6][1=O,2=III,3=,4=,5=,6=]{\text{{#1}\,{\sc {#2}}{#3}{#4}{#5}{#6}}\xspace}
\newcommandx{\semiforbiddenEL}[6][1=O,2=III,3=,4=,5=,6=]{\text{{#1}\,{\sc{#2}}]{#3}{#4}{#5}{#6}}\xspace}
\newcommandx{\forbiddenEL}[6][1=O,2=III,3=,4=,5=,6=]{\text{[{#1}\,{\sc{#2}}]{#3}{#4}{#5}{#6}}\xspace}

\newcommandx{\NVL}[1][1=1243]{\permittedEL[N][v][\textlambda][#1]}
\newcommandx{\NVall}{\permittedEL[N][v][\textlambda][\textlambda][1239,][1243]}

\newcommandx{\CIIall}{\semiforbiddenEL[C][ii][\textlambda][\textlambda][2324--][2329]}

\newcommandx{\NIVL}[1][1=1486]{\semiforbiddenEL[N][iv][\textlambda][#1]}

\newcommandx{\CIVL}[1][1=1550]{\permittedEL[C][iv][\textlambda][#1]}
\newcommand{\CIVall}{\permittedEL[C][iv][\textlambda][\textlambda][1548,][1551]}

\newcommandx{\HeIIL}[1][1=1640]{\permittedEL[He][ii][\textlambda][#1]}

\newcommandx{\semiOIIIL}[1][1=1666]{\semiforbiddenEL[O][iii][\textlambda][#1]}

\newcommandx{\NIIIL}[1][1=1750]{\semiforbiddenEL[N][iii][\textlambda][#1]}

\newcommandx{\CIII}{\semiforbiddenEL[C][iii]}
\newcommandx{\CIIIL}[1][1=1909]{\semiforbiddenEL[C][iii][\textlambda][#1]}

\newcommandx{\NeIVL}[1][1=2424]{\forbiddenEL[Ne][iv][\textlambda][#1]}

\newcommandx{\MgIIL}[1][1=2803]{\permittedEL[Mg][ii][\textlambda][#1]}

\newcommandx{\NeVL}[1][1=3426]{\forbiddenEL[Ne][v][\textlambda][#1]}

\newcommand{\OII}{\forbiddenEL[O][ii]}
\newcommandx{\OIIL}[1][1=3727]{\forbiddenEL[O][ii][\textlambda][#1]}
\newcommand{\OIIall}{\forbiddenEL[O][ii][\textlambda][\textlambda][3726,][3729]}

\newcommandx{\NeIIIL}[1][1=3869]{\forbiddenEL[Ne][iii][\textlambda][#1]}

\newcommand{\OIII}{\forbiddenEL[O][iii]}
\newcommandx{\OIIIL}[1][1=5007]{\forbiddenEL[O][iii][\textlambda][#1]}
\newcommand{\OIIIall}{\forbiddenEL[O][iii][\textlambda][\textlambda][4959,][5007]}

\newcommandx{\NIL}[1][1=5200]{\forbiddenEL[N][i][\textlambda][#1]}

\newcommandx{\OIL}[1][1=6300]{\forbiddenEL[O][i][\textlambda][#1]}

\newcommandx{\HeIL}[1][1=5875]{\permittedEL[He][i][\textlambda][#1]}

\newcommandx{\NIIL}[1][1=6584]{\forbiddenEL[N][ii][\textlambda][#1]}

\newcommand{\SIIall}{\forbiddenEL[S][ii][\textlambda][\textlambda][6716,][6731]}

\newcommandx{\OIIAuL}[1][1=7325]{\forbiddenEL[O][ii][\textlambda][#1]}

\newcommandx{\CIIL}[1][1=158]{\forbiddenEL[C][ii][\textlambda][{#1}\mum]}
\newcommandx{\CIIonly}{\forbiddenEL[C][ii]}

\usepackage[dvipsnames,hyperref]{xcolor}
\usepackage{multirow}
\usepackage[breaklinks=true, colorlinks=true, citecolor=blue, linkcolor=blue,urlcolor=blue]{hyperref}

\hypersetup{
    pdftitle={CR7 Marconcini JWST},
    }

\usepackage[defaultsans]{lato}
\usepackage[LGR, T1]{fontenc}
\newcommand{\msun}{\ensuremath{\mathrm{M_\odot}}\xspace}
\newcommand{\zsun}{\ensuremath{\mathrm{Z_\odot}}\xspace}
\newcommand{\mstar}{\ensuremath{\mathrm{M_\star}}\xspace}
\newcommand{\kms}{\ensuremath{\mathrm{km\,s^{-1}}}\xspace}
\newcommand{\peryr}{\ensuremath{\mathrm{yr^{-1}}}\xspace}
\def\arcsec{$^{\prime\prime}$}
\def\arcmin{$^{\prime}$}

\newcommand{\JWST}{\textit{JWST}\xspace}

\let\oldAA\AA
\renewcommand{\AA}{\text{\oldAA}\xspace}
\let\oldarcsec\arcsec
\renewcommand{\arcsec}{\text{\oldarcsec}\xspace}
\usepackage{mathptmx}
\usepackage{amsmath}
\usepackage[T1]{fontenc}
\DeclareRobustCommand{\VAN}[3]{#2}
\let\VANthebibliography\thebibliography
\def\thebibliography{\DeclareRobustCommand{\VAN}[3]{##3}\VANthebibliography}
\usepackage{graphicx}	
\usepackage{amsmath}	
\begin{document} 
\title{GA-NIFS: Dissecting the multiple sub-structures and probing their complex interactions in the \Lyalpha emitter galaxy CR7 at z = 6.6 with JWST/NIRSpec}
\titlerunning{Dissecting the multiple sub-structures in the \Lyalpha emitter CR7}


\author{
C. Marconcini\inst{1,2,3}\thanks{e-mail: cosimo.marconcini@unifi.it}
\and F.D'Eugenio \inst{1,4}
\and R. Maiolino \inst{1,4,5}
\and S. Arribas \inst{6}
\and A. Bunker \inst{7}
\and S. Carniani \inst{8}
\and S. Charlot \inst{9}
\and M. Perna \inst{6}
\and B. Rodr\'iguez Del Pino \inst{6}
\and H. \"Ubler \inst{10}
\and P. G. P\'erez-Gonz\'alez \inst{6}
\and C. J. Willott \inst{11}
\and T. B\"oker \inst{12}
\and G. Cresci \inst{3}
\and M. Curti \inst{13}
\and I.~Lamperti \inst{2,3}
\and J. Scholtz \inst{1,4}
\and E. Parlanti \inst{8}
\and G. Venturi \inst{8}
}

\institute{
Kavli Institute for Cosmology, University of Cambridge, Madingley Road, Cambridge CB3 0HE, UK
\and
Dipartimento di Fisica e Astronomia, Università degli Studi di Firenze, Via G. Sansone 1,I-50019, Sesto Fiorentino, Firenze, Italy
\and
INAF - Osservatorio Astrofisico di Arcetri, Largo E. Fermi 5, I-50125, Firenze, Italy
\and
Cavendish Laboratory, University of Cambridge, 19 J. J. Thomson Ave., Cambridge CB3 0HE, UK
\and
Department of Physics and Astronomy, University College London, Gower Street, London WC1E 6BT, UK
\and
Centro de Astrobiología (CAB), CSIC-INTA, Ctra. de Ajalvir km 4, Torrejón de Ardoz, E-28850, Madrid, Spain
\and
University of Oxford, Department of Physics, Denys Wilkinson Building, Keble Road, Oxford OX13RH, United Kingdom
\and
Scuola Normale Superiore, Piazza dei Cavalieri 7, I-56126 Pisa, Italy
\and
Sorbonne Université, CNRS, UMR 7095, Institut d’Astrophysique de Paris, 98 bis bd Arago, 75014 Paris, France
\and
Max-Planck-Institut f\"ur extraterrestrische Physik (MPE), Gie{\ss}enbachstra{\ss}e 1, 85748 Garching, Germany
\and
NRC Herzberg, 5071 West Saanich Rd, Victoria, BC V9E 2E7, Canada
\and
European Space Agency, c/o STScI, 3700 San Martin Drive, Baltimore, MD 21218, USA
\and
European Southern Observatory, Karl-Schwarzschild-Strasse 2, 85748 Garching, Germany
}
\authorrunning{C. Marconcini et~al.}

   \date{}

\abstract{
We present JWST/NIRSpec integral field spectroscopic (IFS) observations of the \Lyalpha emitter CR7 at z $\sim$ 6.6, observed as part of the GA-NIFS program. Using low-resolution PRISM (R $\sim$ 100) data, we confirm a bright \Lyalpha emitter, and a diffuse \Lyalpha halo extending up to 3 kpc from the peak of ionized emission, both of them associated to the most massive, UV bright galaxy in the system (CR7-A). We confirm the presence of two additional UV-bright satellites (CR7-B and CR7-C) detected at projected distances of 6.4 and 5.2 kpc from the primary source. We perform SED fitting of the low-resolution data and revealed an inverted star formation history between two satellites at early epochs and a spatially resolved anti-correlation of the gas-phase metallicity and the star formation rate density, likely driven by the gas exchange among the satellites, favouring the merger scenario for CR7. From the high-resolution G395H (R $\sim$ 2700) data, we discover at least three additional companions mainly traced by the \OIIIL emission line, although they are not detected in continuum. We disentangle the kinematics of the system and reveal extended ionised emission linking the main galaxy and the satellites. We spatially resolve the \OIIIL, \OIII[4363], and \Hgamma emission lines and use a diagnostic diagram tailored to high-z systems to reveal tentative evidence of AGN ionisation across the main galaxy (CR7-A) and the N-E companion (CR7-B). Moreover, we detect an unresolved blue-shifted outflow from one of the satellites and present first evidence for a redshifted outflow from the main galaxy. Finally, we compute resolved electron temperature (T$_e \sim 1.6 \times 10^4$ K) and metallicity maps (log(Z/\zsun) from --0.8 to --0.5), and provide insights on how the physical properties of the system evolved at earlier epochs.
}


\maketitle



\section{Introduction}\label{Sec_intro}

The launch of the \textit{James Webb} Space Telescope (\JWST) has established a new era for observational astronomy, revolutionizing our understanding of galaxy properties and assembly processes in the early Universe (redshift \textit{z} $\geq$ 5.5, \textit{t} $\leq$ 1 Gyr). Recent observations of galaxies at the epoch of reionization revealed for the first time their stellar population and age, providing key insights into the star-formation process and star-formation history (SFH) in pristine environments \citep{Whitler2023, Chen2023, Endsley2023, Santini2023, Treu2023, Topping2024, Weibel2024}. The combination of \JWST 's unmatched sensitivity and spectral coverage allow us to trace for the first time the rest-optical emission of galaxies at z$>4$, resolving the internal structure of primordial star forming systems. Moreover, the superb sensitivity of \JWST NIRSpec \citep{Jakobsen2022} in the near-infrared (NIR) wavelength range allow us to investigate the distribution of the ionised gas in primordial galaxies, providing new insights into the interstellar medium (ISM) properties at high-redshift, such as the electron density and gas-phase metallicity \citep[][Cresci et al. in prep]{Taylor2022, Schaerer2022, Rhoads2023, Isobe2023, Curti2023, Curti2024, Abdurrouf2024, Sarkar2024, Marconcini2024}. 

Simulations and theoretical models predict that powerful sources of ionizing photons are crucial to define the physical conditions during the reionization era (z $\sim$ 6-10), i.e. when most hydrogen in the Universe transitioned from the neutral to the ionized phase, driven by energetic UV photons (Lyman continuum or LyC photons, with wavelengths shorter than 912 $\AA$) \citep{Partridge1967, Ciardi2003, Fan2006, Stark2016, Ma2020, Maji2022}. Due to the intervening intergalactic medium (IGM) it is challenging to directly trace LyC photons at high redshift, and thus a valid alternative is to study \Lyalpha emitter (LAE) galaxies, as their properties are fundamental to shed light on the reionisation of intergalactic hydrogen \citep{Blanc2011, Robertson2013, Dijkstra2016, Goovaerts2024}. In particular, the escape fraction of \Lyalpha photons is observed to correlate with the escape fraction of the energetic LyC photons \citep{Dijkstra2016, Xu2022, Izotov2022, Yuan2024}, thus making LAEs the optimal candidates to comprehend the process of cosmic reionisation. Recent \JWST observations played a major role in investigating the physical properties of LAEs during the epoch of reionization, providing stringent constraints on the ionising photon production, the gas-phase metallicity, and the fraction of escaping \Lyalpha photons \citep{Tang2023, Bunker2023, Jones2024, Kumari2024, Saxena2024S, Munoz2024}.

The goal of this paper is to study the gas and stellar population properties in the merger system of CR7 (COSMOS Redshift 7; z = 6.60425). This source was first identified by \cite{Matthee2015} using Suprime-Cam on the Subaru telescope to inspect a region of the COSMOS/UltraVISTA field and detect bright \Lyalpha emitters in the reionization era. \cite{Sobral2015a} performed follow-up spectroscopic observations with X-shooter, SINFONI, and FORS2 at the VLT and DEIMOS at Keck and confirmed CR7 as one of the most luminous \Lyalpha emitters at z $\ge$6 (L$_{\rm \Lyalpha}$ $\sim$ 10$^{44}$ erg s$^{-1}$). Moreover, \cite{Sobral2015a} disfavoured the presence of an Active Galactic Nuclei (AGN) or Wolf-Rayet (WR) stars, as they only detected a narrow \HeIIL emission line and no metal lines in their X-shooter spectrum covering wavelengths from the UV to the NIR. Instead, they proposed a combination of a PopIII-like stellar population ($10^7$ \msun) and an older metal-enriched stellar population ($10^{10}$ \msun, 0.2 \zsun) to explain the mass of the system. \citet{Visbal2016} further investigated this scenario exploring the formation of a massive PopIII cluster through photoionization feedback. They concluded that pristine gas collapsing into the host halo could potentially explain such high PopIII stellar masses in CR7 as a result of a massive PopIII starburst. On the other hand, \citet{Bowler2017} used \textit{Spitzer} photometry in IRAC ch1 and ch2 to provide observational constraints on the \OIIIL line strengths, discarding the presence of PopIII stars in favour of a narrow-line, low-mass AGN or a young, metal-poor starburst.

Further \Lyalpha focused works discussed the possibility that the \Lyalpha luminosity in CR7 might be explained by  a Direct Collapse Black Hole (DCBH) accreting pristine gas \citep{Pallottini2015, Sobral2015a, Dijkstra2016, Hartwig2016, Agarwal2016, Agarwal2017}. In particular, \citet{Pacucci2017} performed radiation-hydrodynamic simulations and found that a 7 $\times$ 10$^6 ~ \msun$ BH accreting metal-free or low-metallicity gas (5 $\times$ 10$^{-3}$ \zsun) is consistent with the DCBH hypothesis and could explain the IR photometry of CR7. 

\begin{figure}
	\includegraphics[width=\linewidth]{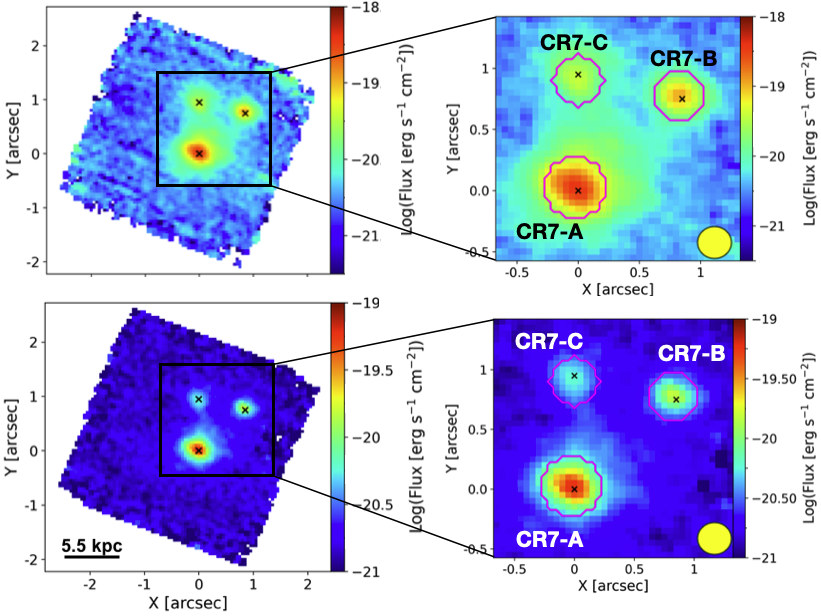}
    \caption{\JWST NIRSpec images from collapsed \OIIIL emission in the high-resolution G395H (top) and low-resolution prism (bottom) data cubes. Sub-cubes with the selected and labelled clumps are shown on the right. The yellow circle is the fiducial PSF of $\sim$ 0.15 \arcsec. Black contours represent the selected regions for the integrated analysis of the three main satellites described in Sec. \ref{Subsec_bo_dust_extinction}. North is up and East to the left.}
    \label{fig:total_images}
\end{figure}

\cite{Matthee2017b_alma} constrained the rest-frame IR continuum of CR7 with band 6 ALMA observations showing the presence of at least four \CIIL[158] satellites, compared to the three observed in the UV \citep{Matthee2015}. They showed that the two brightest satellites are observed in both \CIIL[158] and UV emission, then other smaller satellites detected only in \CIIL[158] emission are observed in the surrounding of the two least massive companions. In this work the three main UV-bright companions are labelled as CR7-A, CR7-B, and CR7-C, following the nomenclature of previous works and as shown in Fig. \ref{fig:total_images}.
From the dynamical mass measurements, \citet{Matthee2017b_alma} found M$_{\rm dyn}$(CR7-A) = 3.9 $\pm$ 1.7 $\times$ 10$^{10}$ \msun, and estimated CR7-A to be the most massive and therefore main or central galaxy of the halo, with indications of a potential ongoing major merger with CR7-C. They used the ALMA observations to constrain the star-formation rate (SFR) in each component, and found values of 28, 5, 7 \msun yr$^{-1}$, for CR7-A, CR7-B, and CR7-C, respectively. Overall, \cite{Matthee2017b_alma} found gas metallicities for CR7-A of Z $\sim$ 0.1 - 0.2 \zsun (i.e., 12 + log(O/H) $\sim$ 7.7 - 8.0), which is entirely inconsistent with the PopIII-like stellar population scenario. Similarly, \citet{Dors2018} found Z $\sim$ 0.1 - 0.5 \zsun fitting UV emission lines with photoionisation models. Finally, \citet{Sobral2019} re-analysed SINFONI data and provided a new, lower, $\sim$ 2 $\sigma$ detection of the \HeIIL emission line. They concluded that all three of CR7-A, B and C are consistent with a young starbursts scenario and a low gas-phase metallicity, in agreement with previous works \citep{Matthee2017b_alma, Dors2018}. Moreover, they argue that based on current data availability there is no evidence for the presence of an AGN in any of the satellites of the system, with only a tentative evidence for CR7-C to have a higher ionization parameter and to potentially host an AGN. This scenario is supported by the comparison of UV emission lines with photoionisation models, which tend to attribute a non-thermal nature to the central source of CR7-C, and thus pointing towards the AGN scenario \citep{Dors2018}.

\cite{Matthee2020} analysed spatially resolved \Lyalpha observations with VLT/MUSE and reported a total \Lyalpha luminosity fainter by a factor of 1.5 with respect to previous works \citep[][]{Sobral2015a}. They observed a $\sim$1 kpc offset between the \Lyalpha and UV peak over CR7-A, with the former being elongated towards CR7-B. Moreover, they explained the observed \Lyalpha equivalent width (EW) with a metal-poor starburst only, without the need to include high-ionisation sources. Finally, by spatially resolving the \Lyalpha morphology they confirmed that the main contribution to the total emission originates from CR7-A, and the elongated nature of the \Lyalpha halo might indicate the presence of multiple star-forming regions, consistent with the presence of three UV clumps, and possibly multiple sub-clumps \citep[e.g.][]{Matthee2017b_alma}.

While this system has been extensively studied  with both ground- and space-based instruments targeting the rest-UV and FIR bands, \JWST/NIRSpec is the first instrument capable of measuring the rest-frame optical spectrum of this intriguing source, to shed light on its nature and provide insights on galaxy assembly processes via mergers at high redshift. 
In this work, we present spatially resolved high resolution (R $\sim$ 2700) and low resolution (R $\sim$ 100) NIRSpec/IFS observations of CR7, providing, for the first time, spatially resolved rest-frame optical spectroscopy of this intriguing system.  This paper is organised as follows. In Sec. \ref{Sec2_observations} we present the NIRSpec observations, together with the data reduction. In Sec. \ref{Sec3_analysis_line_fitting_line_ratios} we discuss the emission line analysis of the G395H and prism data cubes, the emission line fitting method and the following results via line ratio analysis and spatially resolved kinematics. Then, in Sec. \ref{Sec4_discussion} we discuss the implications of our results in the broad context of galaxy assembly processes. Finally, in Sec. \ref{Sec5_summary_conclusion} we summarize the main results, comparing our findings with previous works. In this paper, we assume the \citet{Planck2020} cosmology, i.e. a flat $\Lambda$CDM cosmology with H$_0$= 67.4 km s$^{-1}$ Mpc$^{-1}$, $\Omega_\mathrm{m}$ = 0.315, and $\Omega_\Lambda$ = 0.685.

\section{Observations and data reduction}\label{Sec2_observations}

\begin{figure*}
\includegraphics[width=\linewidth]{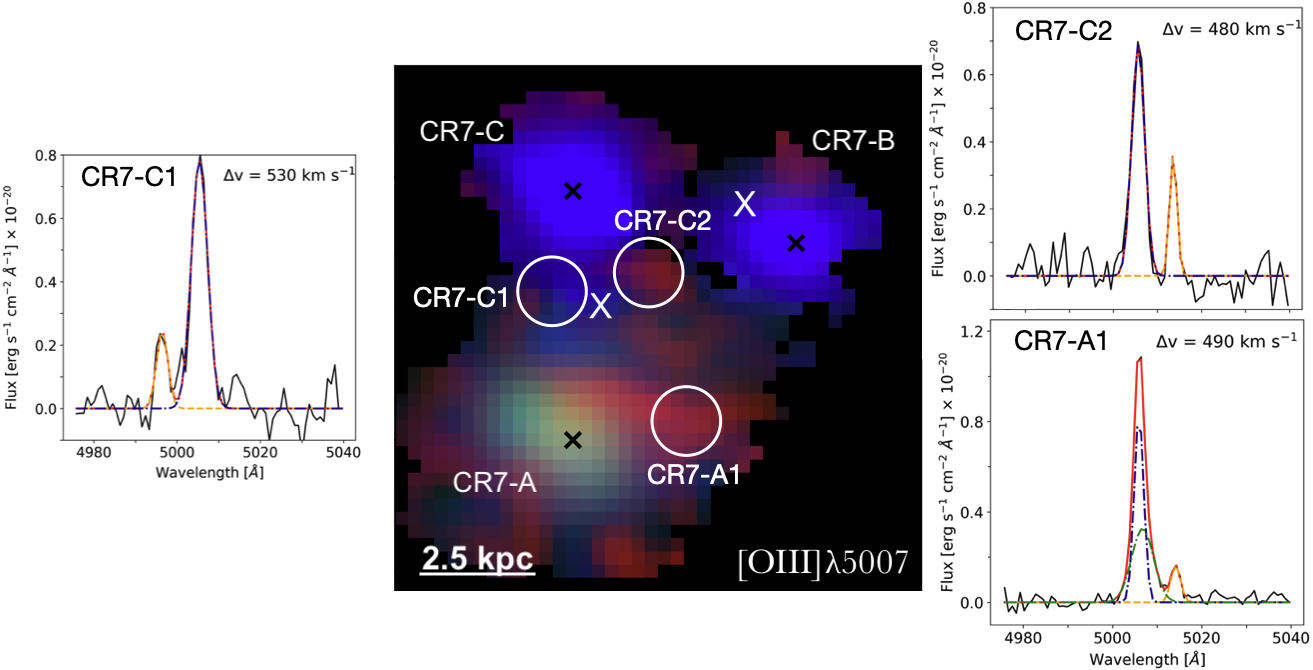}
    \caption{Three-color emission line image of CR7 derived from the G395H data cube. The emission line maps are obtained by integrating the continuum-subtracted data-cube around the \OIIIL emission line in the velocity ranges v $\leq$ -- 200 km s$^{-1}$ (blue), -- 200 km s$^{-1}$ $\leq$ v $\leq$ 200 km s$^{-1}$ (green) and v $\ge$ 200 km s$^{-1}$ (red), with respect to the systemic redshift z = 6.60425 (see Sec. \ref{Subsec3_kinematic}). The black crosses mark the positions of the three main components shown in Fig. \ref{fig:total_images}. White crosses mark the positions of the two brightest additional \CIIL components detected by \citet{Matthee2017b_alma}. Solid white circles and corresponding labels mark the putative location of unresolved, kinematically distinct sub-satellites, identified by the detached orange component (see Sec. \ref{Subsec3_kinematic}). The small panels show the integrated spectra and best-fit of the \OIIIL emission of such sub-satellites. Data are in black, the total best fit model is red. The narrow, broad and detached Gaussian components are in blue, green, and orange, respectively. The velocity difference between the peak of the narrow and detached components is shown in each small panel. See Sec. \ref{Narrow_broad_section} for details on the components definition.}
    \label{fig:rgb_image}
\end{figure*}

\subsection{NIRSpec Observations}\label{Sec2_nirspec_observations}
CR7 (CR7-COS-6.60-L, R.A.  10$\rm ^h$ 00$\rm ^m$ 58.005$\rm ^s$ Dec. +01$^{\circ}$ 48\arcmin 15.25\arcsec, J2000) was observed as part of the Galaxy Assembly with NIRSpec Integral Field Spectroscopy (GA-NIFS\footnote{GA-NIFS website: \url{https://ga-nifs.github.io}}) Guaranteed Time Observations (GTO) programme, included in proposal $\#$1217 (PI: N. Luetzgendorf). The observations were carried out on May 2 2023 with the NIRSpec IFU \citep[field of view, FOV: 3.1\arcsec $\times$ 3.2\arcsec, spaxel size: 0.1\arcsec;][]{Boker2022, Rigby2023}. The observations include both high-(R$\sim$2700) and low-(R$\sim$100) resolution configurations with the G395H grating and the PRISM, respectively, with a 0.5\arcsec size cycling pattern of eight dithers, covering a total effective FOV of about 3.7\arcsec$\times$3.7\arcsec. The total integration times were 18207 seconds (5h) for G395H and 3968 seconds (1.1h) for the PRISM configuration.

We used the \OIIIL and \Hbeta integrated emission line profiles over the aperture of CR7-A to estimate the redshift, assuming CR7-A to be the main galaxy of the system. In the following we will adopt z = 6.60425 $\pm$ 0.00002  as the systemic redshift of the system, thus defining the zero velocity of the gas kinematics.

\subsection{Data Reduction}\label{Sec2_data_reduction}
We performed the data reduction using the \JWST calibration pipeline version 1.8.2 with CRDS context jwst1068.pmap. The steps and changes we performed with respect to the standard pipeline in order to improve the final data quality are discussed in \cite{perna2023_pipeline}. To combine each integration and create the final data cube with spaxel size of 0.05\arcsec, we adopted the $drizzle$ method, using an official patch to account for a known issue affecting the calibration pipeline\footnote{\url{https://github.com/spacetelescope/jwst/pull/7306}}. For the high-resolution data, we performed no background subtraction since it is accounted for during the emission line fitting. For the low resolution PRISM cube, we instead performed a background subtraction using selected regions away from the three main clumps.
\section{Analysis and Results}\label{Sec3_analysis_line_fitting_line_ratios}
\subsection{High-resolution data - Emission line fitting}\label{Subsec3_line_fitting}
The G395H high-resolution data cube was analysed by means of a set of custom Python scripts in order to subtract the continuum emission and then fit the observed emission lines. As a first step, we performed a Voronoi tessellation binning \citep{Cappellari2003} on the continuum level between 3800 and 6900 \AA rest-frame, requiring  a minimum signal-to-noise ratio (SN) in each spectral channel of 3. Spaxels with SN $\geq$ 3 are not binned. Then we used the Penalized Pixel-Fitting software \citep[\textsc{pPXF};][]{Cappellari2004, Cappellari2023} to fit the continuum in each bin. We used single-stellar population templates from the stellar-population synthesis tool \textsc{fsps} \citep{Conroy2009, Conroy2010}, combining them with a third order multiplicative polynomial. While fitting the continuum, we simultaneously fitted the emission lines with one or two Gaussian components to account for possible absorptions underlying the emission lines. The criteria adopted to decide the Gaussian components, as well as other fitting details, are described in the following. Bad pixels, as well as the detector gap are masked during this step in order to maximise the fit quality. Then, we selected and subtracted the continuum in each bin from the corresponding flux of each spaxel, obtaining a continuum-subtracted data cube which will be used to infer the gas properties via a detailed multi-Gaussian emission line fitting. We performed a spatial smoothing to match the spatial resolution within the FOV by convolving each spectral slice of the data cube with a $\sigma$ = 1 spaxel (i.e., 0.05\arcsec) Gaussian kernel. To extract the emission-line properties we fit the emission lines of the continuum-subtracted data cube with up to three Gaussians, tying the velocity and the velocity dispersion of each Gaussian component among all the emission lines. After a detailed inspection of the high-resolution data cube in each spaxel we imposed reasonable boundaries on the Gaussian components, in order to reproduce the observed asymmetric line profiles. As an example, Fig. \ref{fig:rgb_image} shows a three colour emission line image of CR7, highlighting the complexity of the \OIIIL emission line in three distinct unresolved regions, from now on labelled as CR7-C1, CR7-C2, and CR7-A1. The spectral profile of \OIIIL across these three regions is shown in Fig. \ref{fig:rgb_image}, with the presence of blue-shifted (CR7-C1) and red-shifted (CR7-C2 and CR7-A1) components, shown as partially or completely detached peaks. To account for these features we adopted tailored boundaries on the velocity and velocity dispersion of each Gaussian component. In particular, the first component is used to model the brightest, narrowest and closest to systemic velocity emission line profile (blue dashed-dot, labelled as \textit{narrow} component). The second component is used to reproduce the possible broad line profile (green dashed-dot, labelled as \textit{broad} component), and finally we used a third low-velocity dispersion, high projected velocity component (yellow dashed line, labelled as \textit{detached} component). A detailed discussion of different kinematic components is presented in Sec. \ref{Narrow_broad_section}.

The intensities for all the components are free to vary. We accounted for the \OIIIall emission line doublet by fixing the flux ratio to 1/3 between the two lines, given by the Einstein coefficients of the two transitions \citep{Osterbrock2006}. We decided the optimal number of Gaussian components to be used in each spaxel based on the results of a Kolgomorov-Smirnov test \citep[see][for details]{Marasco2020}. In particular, we compared the residuals of the fit with a different number of Gaussians in the wavelength range between 4980 and 5040 \AA rest-frame, in order to optimise the fit result for the asymmetric \OIIIL emission line. If the residuals of the n-component Gaussian fit statistically improve the fit with respect to the one-component Gaussian fit then we adopt n Gaussians to build the final best-fit model. This procedure gives the minimum and optimal number of Gaussian components to be used to reproduce the observed emission line profile. As a result, we obtain a two-dimensional map with the optimal number of Gaussian components to be used in each spaxel, highlighting structured regions that benefit from multiple components (up to 3) to reproduce the asymmetric or broad line profile, ultimately providing insightful information on the intrinsic gas kinematics. The final result is an emission-line-only model cube centred around each emission line. 

\begin{figure*}\includegraphics[width=\linewidth]{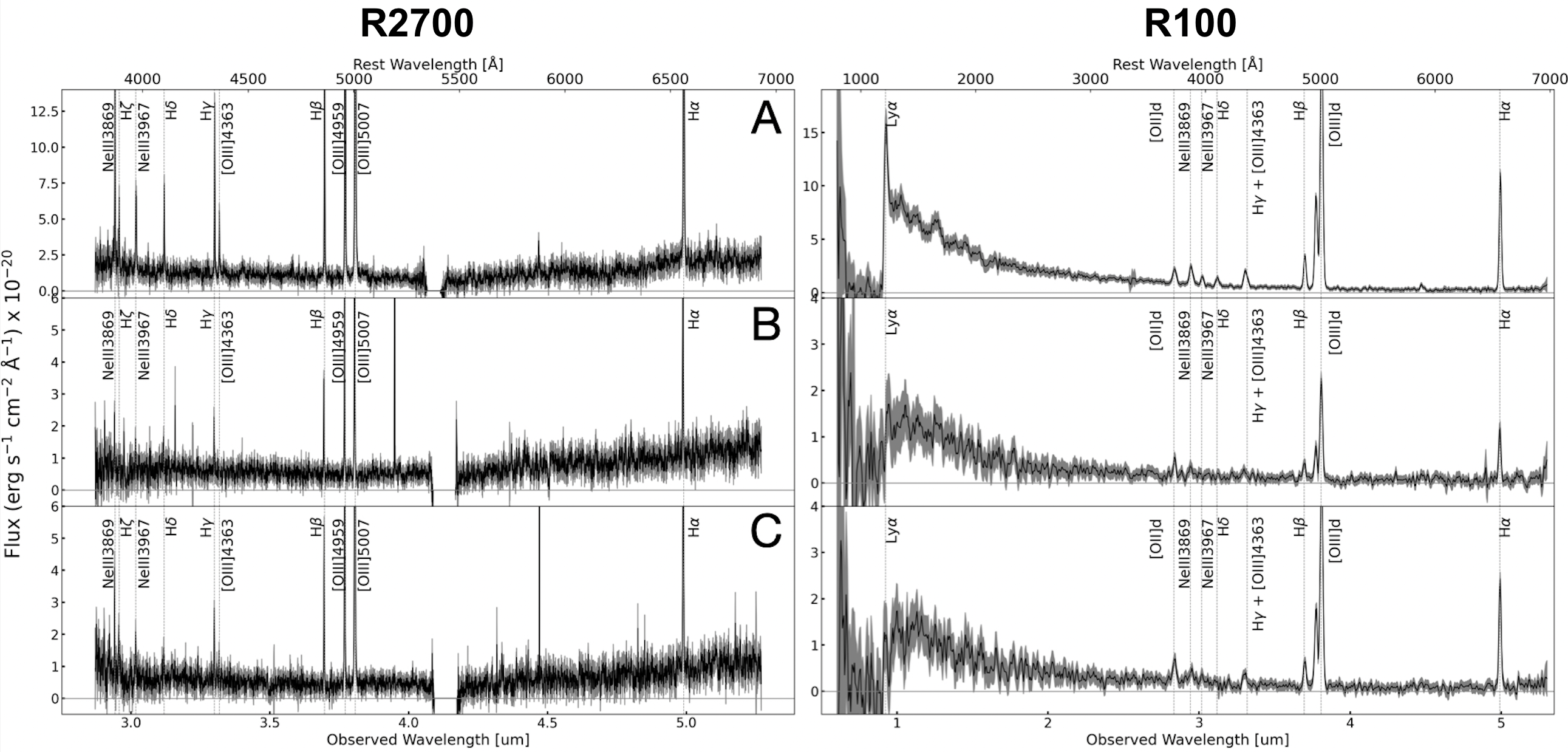}
    \caption{JWST/NIRSpec G395H (left) and prism (right) integrated spectra from CR7-A, CR7-B, and CR7-C, from top to bottom panels, respectively. The high-resolution spectra cover the wavelength range from 2.7 to 5.27 $\mu$m. The low resolution, background-subtracted spectra show prominent \Lyalpha emission line in CR7-A.}
    \label{fig:spettri_grism_prism}
\end{figure*}

The top panels in Fig.~\ref{fig:total_images} show images of the total and selected sub-region of the G395H data cube obtained collapsing over the total \OIIIL emission. The bottom panels show the same \OIIIL emission extracted from the prism data cube. 
The left and right panels in Fig.~\ref{fig:spettri_grism_prism} show the integrated spectra of the high- and low-resolution (background subtracted) spectra extracted from the three apertures shown in Fig. \ref{fig:total_images}, corresponding to CR7-A, CR7-B, and CR7-C. As shown by the integrated low-resolution spectra in Fig.~\ref{fig:spettri_grism_prism} we significantly detect the \Lyalpha emission line only in CR7-A, with no evidence in CR7-B and CR7-C which instead show evidence of damped \Lyalpha absorption. We will further discuss this feature in Sec. \ref{Subsec_prism_line_fitting}. 
In this work, we focused on a `subcube' centred on the main galaxy and shown in the right panels of Fig.~\ref{fig:total_images}, since no additional source is detected outside of the FOV of the selected subcube. Finally, in the high resolution data cube we fitted the following emission lines: \Halpha, \Hbeta, \Hgamma, \Hdelta, \Hepsilon, \Hzeta, \HeIL[3889], \HeIL[5876], \NeIIIL, \NeIIIL[3967], \OIIIL[4363], \OIIIL[4959], and \OIIIL. Table \ref{tab:line_fluxes} shows the integrated fluxes from the apertures shown in Fig. \ref{fig:total_images} for the detected emission lines in the CR7 system.
\begin{table}
    \caption{Integrated emission line fluxes from the apertures shown in Fig. \ref{fig:total_images} for CR7-A, CR7-B, and CR7-C. All values listed are derived from the R2700 data cube except for the \HeIIL, \semiOIIIL, and \OII[3727], which are derived from the R100 data cube and represent a 1 $\sigma$ upper limit, if the error is not reported. \OII[3727] is the blend of the \OIIall doublet. Spaxels with SN $\leq$ 3 are masked. Fluxes are in units of 10$^{-18}$ erg s$^{-1}$ cm$^{-2}$.}
    \centering
    \begin{tabular}{cccc}
        \hline
        Line ID & CR7-A & CR7-B & CR7-C \\
        \hline
        \OIIIL & 62.6 $\pm$ 0.4 & 12.5 $\pm$ 0.3 & 5.8 $\pm$ 0.3 \\
        \Halpha & 21.5 $\pm$ 0.7 & 4.3 $\pm$ 0.5 & 2.3 $\pm$ 0.5 \\
        \Hbeta & 7.3 $\pm$ 0.4 & 1.5 $\pm$ 0.3 & 0.9 $\pm$ 0.3 \\
        \Hgamma & 3.2 $\pm$ 0.4 & 0.3 $\pm$ 0.2 & -- \\
        \Hdelta & 1.8 $\pm$ 0.5 & 0.3 $\pm$ 0.2 & 0.2 $\pm$ 0.2 \\
        \Hepsilon & 1.5 $\pm$ 0.4 & 0.2 $\pm$ 0.2 & -- \\
        \Hzeta & 0.2 $\pm$ 0.2 & 0.1 $\pm$ 0.1 & -- \\
        \HeIL[3889] & 0.8 $\pm$ 0.4 & 0.1 $\pm$ 0.1 & -- \\
        \NeIIIL & 4.6 $\pm$ 0.7 & 0.8 $\pm$ 0.3 & 0.4 $\pm$ 0.3 \\
        \NeIIIL[3967] & 2.3 $\pm$ 0.5 & 0.3 $\pm$ 0.2 & -- \\
        \OIIIL[4363] & 1.15 $\pm$ 0.4 & 0.1 $\pm$ 0.1 & -- \\
        \HeIL[5876] & 0.8 $\pm$ 0.3 & -- & -- \\
        \HeIIL & $\leq$0.3 & -- & -- \\
        \semiOIIIL & $\leq$0.4 & -- & -- \\
        \OII[3727] & 5.5 $\pm$ 1.6 & 1.8 $\pm$ 0.4 & 1.4 $\pm$ 0.4 \\
        \hline
    \end{tabular}
    \label{tab:line_fluxes}
\end{table}

\subsection{Dust attenuation}\label{Subsec_bo_dust_extinction}
To estimate the dust attenuation we used the \Halpha/\Hbeta line ratio and assumed a Calzetti extinction law \citep[][]{Calzetti2000} to measure A$_V$ and to correct the observed fluxes of each line for dust attenuation\footnote{For completeness, we also estimated the dust attenuation assuming a Cardelli extinction law \citep{Cardelli89}. Nevertheless, we observed that adopting such attenuation law only varies the resulting extinction by $\leq$ 10\%, which does not alter our conclusions.}. 
We performed an unresolved analysis for the dust attenuation by integrating the spectra from the apertures over the three components, where both \Halpha and \Hbeta have SN larger than 5. We estimated a nebular attenuation of A$_V$ = 0.18$\pm$0.08, 0.04$\pm$0.02, 0.07$\pm$0.03 for CR7-A, CR7-B, and CR7-C, respectively. Such low values for A$_V$ are consistent with the previous upper limits for the IR continuum derived from deep ALMA observations. Indeed, \cite{Matthee2017b_alma} estimated dust masses $\leq$ 8$\times$10$^{6}$ \msun based on the FIR continuum, which, assuming dust is enclosed within the radius of the CR7 system ($\sim$ 3 kpc) and a dust opacity $\kappa$=10$^{3}$ cm$^{2}$ g$^{-1}$, provides an attenuation of A$_V$ $\sim$ 0.07 . In the following analysis, line ratios are corrected for dust attenuation if not specified otherwise.
\begin{figure*}\includegraphics[width=\linewidth]{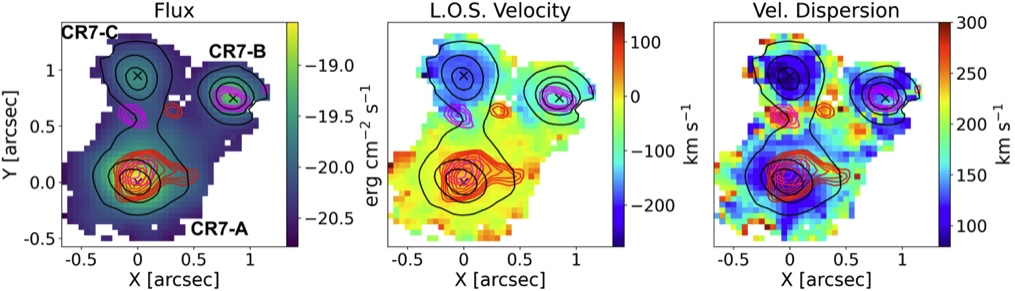}
    \caption{\OIIIL moment maps from G395H data. From left to right, the integrated emission line flux in logarithmic scale, the flux-weighted LOS velocity and the velocity dispersion maps. All moment maps are calculated from the total best fit model. Contours highlight the blue-shifted emission v $\leq$ -- 400 km s$^{-1}$ (magenta),  -- 400 km s$^{-1}$ $\leq$ v $\leq$ 100 km s$^{-1}$ (black) and redshifted v $\geq$ 400 km s$^{-1}$ (red), with respect to the systemic velocity of CR7-A defined by the redshift z = 6.60425 (see Sec. \ref{Subsec3_kinematic}). Moment maps are masked at SN smaller than 3. The black crosses mark the positions of the three main components, labelled in the left panel for clarity. North is up, East to the left.}
    \label{fig:mom_mapso3}
\end{figure*}
\subsection{Gas kinematics}\label{Subsec3_kinematic}
As a result of the spaxel-by-spaxel fit described in the previous section, we obtained the one-, two- or three-Gaussian best fit model for each line in each spaxel and a model cube for each emission line. From this, we can create spatially resolved moment maps and velocity channel maps for any detected emission line, which provide precious hints on the projected gas kinematics. As an example, Fig. \ref{fig:mom_mapso3} shows the integrated emission line flux (moment 0), the flux-weighted line-of-sight (LOS) velocity (moment 1) and the flux-weighted velocity dispersion  (moment 2) maps for the total best fit \OIIIL emission line model. 

The moment maps clearly highlight the clumpy morphology of the system, with the \OIIIL emission peaking at the three well-known UV-bright components \citep{Matthee2015, Matthee2017a, Matthee2017b_alma, Sobral2015a}. From the profile of integrated \OIIIL emission line (Fig. \ref{fig:spettri_grism_prism}) we measure redshifts of z = 6.60425, z = 6.60115, z = 6.59948, for CR7-A, CR7-B, and CR7-C, respectively. Therefore, assuming the redshift of CR7-A as a reference, CR7-B and CR7-C are blue-shifted by about --100 and --200 km s$^{-1}$ with respect to the main galaxy, respectively. 
The integrated emission line flux map in Fig. \ref{fig:mom_mapso3} clearly shows the presence of extended ionised emission between CR7-A and CR7-C with average relative projected velocity of 100 km s$^{-1}$, suggesting a possible interaction among these two components. In addition, extended red-shifted emission is also detected from CR7-A towards the West. Interestingly, the ionised gas morphology shown in Fig. \ref{fig:mom_mapso3} shows fainter diffused ionised emission between CR7-A and CR7-B than between CR7-A and CR7-C.

Different velocity contours are superimposed in Fig. \ref{fig:mom_mapso3} on the \OIIIL moment maps. Red contours represent arbitrary levels of the ionised emission at velocities $\geq$ 400 km s$^{-1}$. Magenta and black contours represent arbitrary levels of the flux at velocities $\leq$ --400 km s$^{-1}$ and --400 km s$^{-1}$ $\leq$ v $\leq$ --100 km s$^{-1}$, respectively. As expected from the integrated spectra shown in Fig. \ref{fig:rgb_image}, we observe clear sub-structures in CR7 at blue-shifted and red-shifted velocities up to 500 km s$^{-1}$. Magenta contours trace the presence of CR7-C1, which is at $\sim$ 1.5 kpc (projected) from CR7-C and located along the blue-shifted elongated emission connecting CR7-A and CR7-C. Moreover, magenta contours also show blue-shifted emission slightly offset from the position of the centres of CR7-A and CR7-B, which likely indicates the presence of outflows from the central source or tidal features. On the other hand, red contours clearly show the presence of CR7-C2, which appears to be completely detached from the other components, both spatially and kinematically. Interestingly, we observe higher velocity dispersion corresponding to the location of CR7-A1, CR7-C1, and CR7-C2 as due to the overlap along the line of sight of multiple kinematic components (as shown in side panels in Fig. \ref{fig:rgb_image}). Extended ionised red-shifted emission connecting CR7-A and CR7-A1 suggests close interaction among these two components. Finally, red contours in Fig. \ref{fig:mom_mapso3} also show elongated emission from the centre of CR7-A towards the South-East, with an almost symmetric shape with respect to the elongated emission extending towards CR7-A1. The detection of clearly distinct peaks of ionised emission at such high projected velocities, with a tentative 2.5 $\sigma$ detection of CR7-A1 also in \Halpha, points towards the presence of multiple, smaller components in CR7, similar to the findings of \citet{Matthee2017b_alma} from \CIIL.

From the spatially resolved \OIIIL emission-line fit we measured an average velocity dispersion\footnote{This includes the contribution from the instrument spectral dispersion.} of CR7-A of 124 $\pm$ 17 km s$^{-1}$.
To calculate a reference dynamical mass, we assumed the structure of CR7-A to be consistent with a disc and used the stellar virial estimator of \citet{vanderWel2022}, using our velocity dispersion and effective radius of 920$\pm$ 60 pc\footnote{We computed the effective radius estimating the distance from the peak emission of CR7-A enclosing 50\% of the ionised emission. The uncertainty is computed performing a bootstrap with 10$^{3}$ iteration.}. Then, assuming a S\'{e}rsic index of 1 and an axis ratio of 0.5, we obtained a dynamical mass of M$_{\rm dyn}$ = 2.4$^{+1.5}_{-0.9}$ $\times$ 10$^{10}$\msun, which is consistent with the previous \CIIonly-based estimate of 3.9 $\pm$ 1.7 $\times$ 10$^{10}$~\msun from \cite{Matthee2017b_alma}. Nevertheless, to compute the dynamical mass of CR7-A we rely on highly uncertain estimates and assume the morphology of CR7-A to be consistent with a disc, for which there is no convincing evidence. Therefore, the value provided here for the dynamical mass has to be considered as order of magnitude estimate. Nevertheless, when using
similar assumptions, the \OIIIL and \CIIL dynamical mass estimates agree very well, showing no evidence of a bias between these two tracers.
As anticipated in the previous section, to further investigate the nature of such extended emissions and possibly detached smaller components, we adopted tailored criteria to discriminate between different kinematic structures, as discussed in the next section.

\begin{figure*}
	\includegraphics[width=\linewidth]{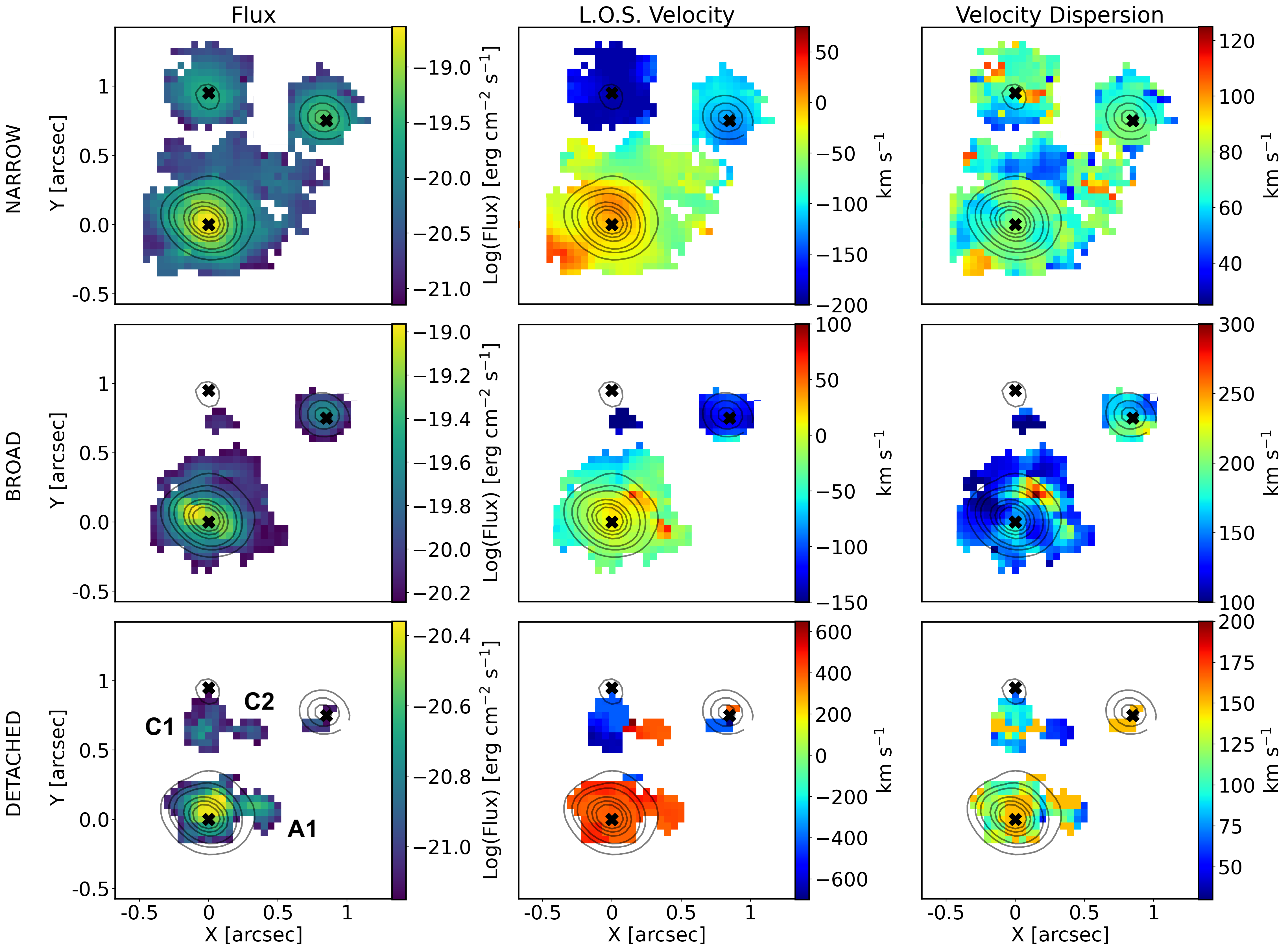}
    \caption{Moment maps of the ionized gas in CR7, traced by the \OIIIL emission line, for three disentangled kinematic components. From left to right: the integrated emission, the line of sight velocity and velocity dispersion maps. From top to bottom: moments maps for the \textit{narrow}, \textit{broad}, and \textit{detached} components (see Sec.\ref{Subsec3_kinematic} for the method used to discriminate  between different components). The first order momentum is computed with respect to z = 6.60425. A SN cut of 3 on the flux of each component was applied to the maps of the \textit{narrow} and \textit{broad} components. A SN cut of 1.5 was applied to the  \textit{detached} component maps. The black crosses mark the positions of the three main components of CR7. Labels for the three sub-components, i.e. CR7-A1, CR7-C1, and CR7-C2 are reported on the bottom left map for clarity.  Black contours are arbitrary \OIIIL flux levels. Moment maps are not corrected for dust attenuation.}
    \label{fig:narrow_broad_detached}
\end{figure*}

\subsubsection{Narrow and Broad components}\label{Narrow_broad_section}
Employing the multi-Gaussians fit described in Section \ref{Subsec3_line_fitting} we were able to reproduce the complex emission line profile in each spaxel. In particular, depending on the location within the system, we observed up to three different kinematic components. First, we observed a narrow emission line profile (most commonly found for CR7-C) with no evident asymmetries or broad wings  which we attribute to systemic emission from the galaxy. Then, we observed a combination of narrow and broad components around CR7-A, with the two components having similar low projected velocities but different line widths. To reproduce such profiles we adopted narrow + broad Gaussian components. Finally, we detected two partially or completely detached narrow Gaussian components with a velocity shift of the fainter component with respect to the brightest component of up to $\pm$ 500 km s$^{-1}$, as shown in Fig. \ref{fig:rgb_image}. As discussed in Sec. \ref{Subsec3_line_fitting}, we refer to these three components as \textit{narrow}, \textit{broad}, and \textit{detached}. Fig. \ref{fig:narrow_broad_detached} shows the three moments maps for the \textit{narrow} (top), \textit{broad} (middle), and \textit{detached} (bottom) components. The separation of the observed line profiles into three kinematic components allows us to better investigate the observed complex ionised gas features. In particular, the \textit{narrow} component traces gas with low projected velocity and velocity dispersion (see Sec. \ref{Subsec3_kinematic} for details). Indeed, the \textit{narrow} component maps represent the systemic emission emerging from each component of CR7, with the integrated flux map showing no sub-structures and a symmetric morphology.

For CR7-A, the main source, the \textit{narrow} component velocity map shows a tenuous north-south  gradient, with amplitude of $\sim$ 20 km s$^{-1}$, suggestive of a system observed close to face-on, under the assumption of a rotating disc. Interestingly, in CR7-A the velocity dispersion is larger along the axis of maximum velocity variation (instead of orthogonally, as expected from beam smearing of a rotating disc). The broad component maps show a peak of emission slightly shifted towards the North-East from the peak of the narrow component, with an elongated structure directed NE-SW, possibly indicating merger or outflow activity. The broad component LOS velocity map shows an almost constant projected velocity over CR7-A with amplitude of --30 $\pm$ 15 km s$^{-1}$. Interestingly, the velocity dispersion map of the broad component shows the maximum values perpendicular to its flux orientation, directed North-South and peaking over a region of high velocity dispersion ($\sigma \geq$ 200 km s$^{-1}$). The peak of velocity dispersion towards North-West of CR7-A has to be ascribed to the broadening of the emission line profile which is leading to the completely detached component CR7-A1 (see Fig. \ref{fig:rgb_image}), as suggested by the co-spatiality with the red contours in Fig. \ref{fig:mom_mapso3}. Before the two components of CR7-A and CR7-A1 separate, the line profile broadens and shifts towards redder projected velocities, leading to the red-shifted region towards North-West of CR7-A. Finally, the \textit{detached} component maps shows a red-shifted faint region surrounding CR7-A, peaking over CR7-A1. This component extends from the North side of CR7-A, following the extended emission and leading to CR7-A1, indicating clear gas flows among these two components.

The narrow moment map shows a blue-shifted peak of emission over CR7-B (with respect to the zero velocity assumed from integrated \OIIIL line profile over CR7-A). The velocity dispersion of the narrow component shows an average constant line broadening with velocity dispersion of 70 $\pm$ 5 km s$^{-1}$. As with CR7-A, we found that a spatially extended broad component is necessary to reproduce the line profile over CR7-B. In particular, we found that the broad component has the same systemic velocity as the narrow component in CR7-B. Moreover, we observed an enhanced velocity dispersion region southward of CR7-B with peaks of 210 km s$^{-1}$, which is almost unresolved. 
Additionally, we found that the same region benefits from the inclusion of a \textit{detached} component, to reproduce the blue-shifted emission. Overall, this region southward of CR7-B appears to have high velocity dispersion as traced both by the \textit{broad} and \textit{detached} ($\sigma$ = 150 km s$^{-1}$) components. This region also appears to be co-spatial with the blue-shifted contours shown in Fig. \ref{fig:mom_mapso3} and with the enhanced \OIIIL/\Hbeta ratio shown in Fig. \ref{fig:O3_over_Hbeta}, suggesting the presence of an unresolved high-ionisation source and outflow/merger scenario.

Finally, we observe that the centre of CR7-C shows a narrow profile with blue-shifted velocity of $\sim$ 200 km s$^{-1}$ with respect to the systemic velocity assumed. Interestingly, we observe that the emission of the blue-shifted contours shown in Fig. \ref{fig:mom_mapso3} is well traced by the \textit{detached} component, tracing the elongated blue-shifted emission extending between CR7-C and CR7-A and culminating in CR7-C1. Similarly, the hypothetical third component CR7-C2 emerges from the \textit{detached} moment maps as an isolated red-shifted low-velocity dispersion region ($\sigma$ $\sim$ 50 km s$^{-1}$).

Overall, from the ionised gas kinematics we find that faint multiple sub-components exist in CR7, as traced by detached narrow line profiles in \OIIIL (Figs.~\ref{fig:rgb_image} and~\ref{fig:narrow_broad_detached}) and as already suggested by previous works \citep{Matthee2017b_alma}. Interestingly, the detached narrow component tracing the emission of CR7-A1 is also detected in \Halpha, even thought its detection is confirmed only at 2.5 $\sigma$. The \textit{detached} component might also be tracing gas flows among such structures, as suggested by the red-shifted elongated structures observed between CR7-A and CR7-A1 and the blue-shifted structure between CR7-C and CR7-C1 (see Fig. \ref{fig:mom_mapso3} and bottom panels Fig. \ref{fig:narrow_broad_detached}). As anticipated, CR7-C2 appears not to be connected to any of the other main- or sub-components in CR7. The \textit{broad} components reasonably trace outflowing or tidal features, mostly from CR7-A towards the North-East and the South-West and from CR7-B southwards (see middle panels in Fig. \ref{fig:narrow_broad_detached}).

\subsection{Excitation Diagnostics and Line Ratios}\label{Subsec3_line_ratios}
In this section we used the results of the emission line fitting of the high-resolution data cube performed in Sec. \ref{Subsec3_line_fitting} to investigate the main line ratios and examine the excitation source in CR7. 

\begin{figure}
	\includegraphics[width=\linewidth]{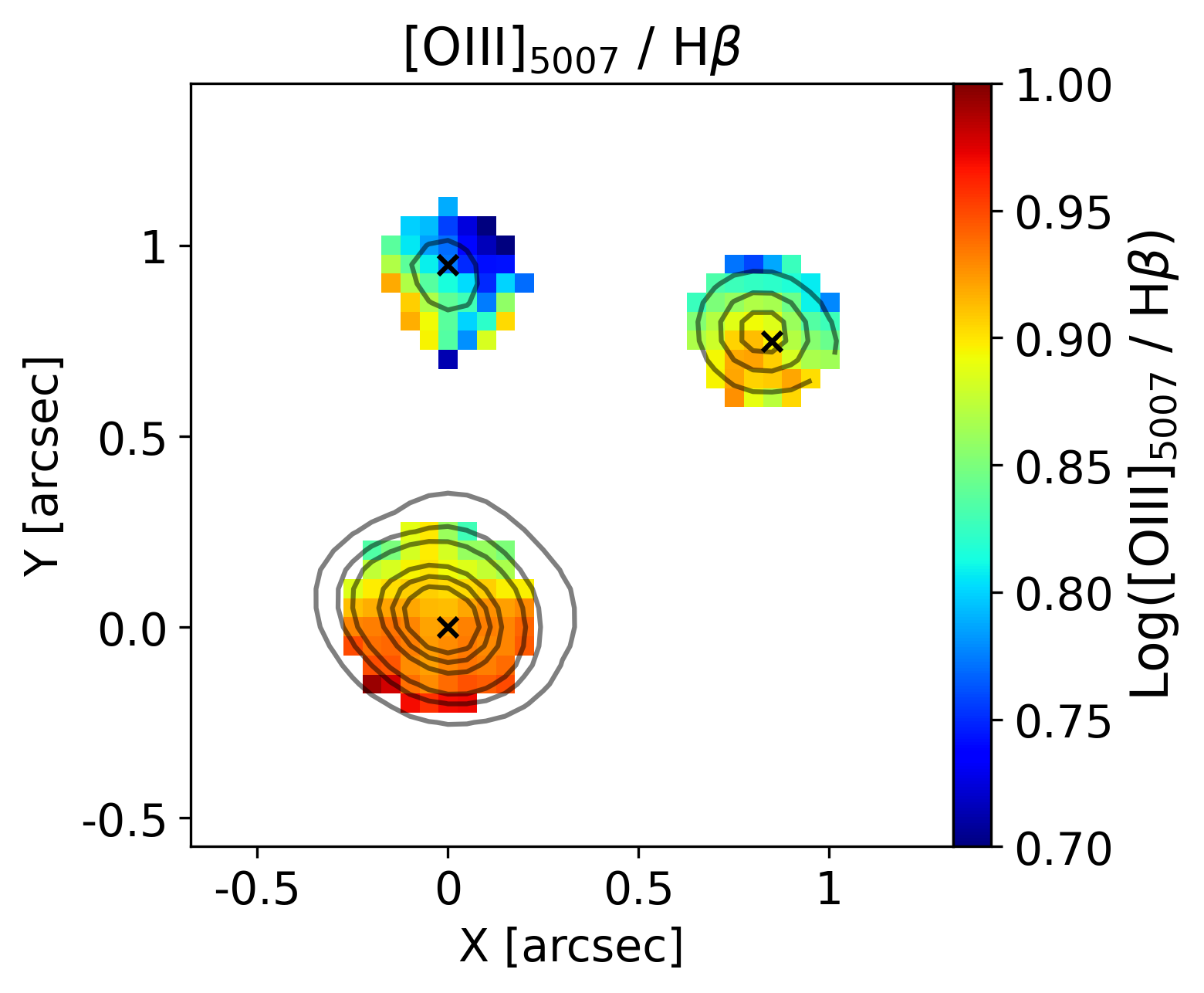}
    \caption{\OIIIL/\Hbeta emission line ratio in CR7 in logarithmic scale. Black solid lines are arbitrary \OIIIL flux levels. Spaxels with SN $\le$ 5 are masked. Emission lines are corrected for dust attenuation.}
    \label{fig:O3_over_Hbeta}
\end{figure}

Recent works highlighted the difficulty in distinguishing between SF and AGN ionisation at high-z \citep[][]{Feltre2016, nakajima2022, Kocevski2023, Ubler2023, Scholtz2023, Maiolino2023AGN_JADES}, employing the standard BPT diagram \citep{Baldwin1981}. Indeed, we do not detect \NIIL in any of the galaxies in the CR7 system, while \OIIIL/\Hbeta generally higher than 5 (Fig. \ref{fig:O3_over_Hbeta}) means that the BPT is inconclusive.

Therefore, we explored the possible source of ionisation in CR7 by employing new  diagnostic diagrams. Specifically, we adopt the diagnostic diagram from \cite{Mazzolari2024}, which 
leverage the \OIIIL[4363] auroral line (when detected) and which
is particularly well suited to identify narrow line TypeII AGN at high redshift.

\begin{figure}
	\includegraphics[width=\linewidth]{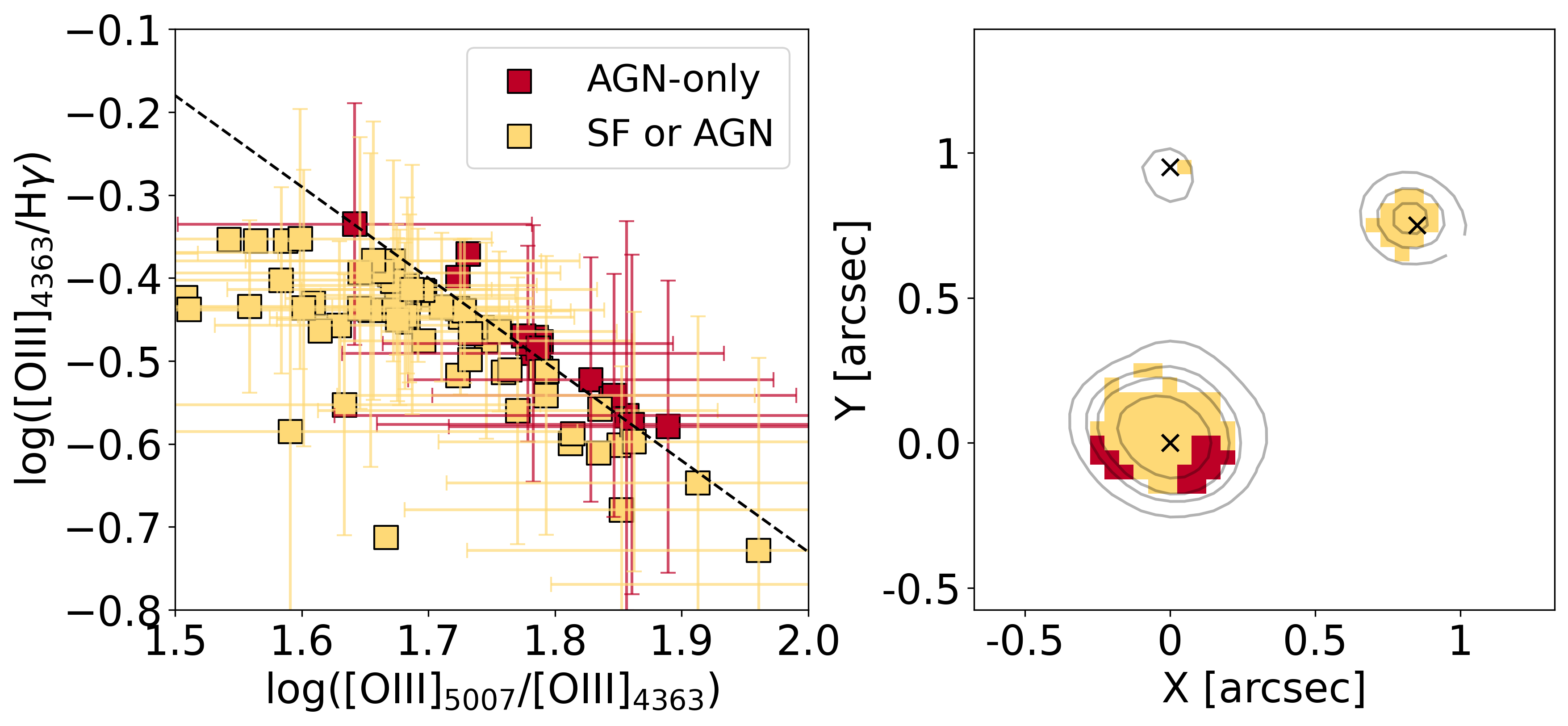}
    \caption{Left: Diagnostic diagram based on \citet{Mazzolari2024} for the excitation source in CR7 based on the \OIIIL[4363]/\Hgamma vs \OIIIL/\OIIIL[4363] line ratio. Dashed black line is the discriminator between AGN or SF excitation source, respectively. Right: Spatially resolved map of CR7 with spaxels color-coded based on their position on the diagram on the left. For visualisation clarity we reported error-bars only for half of the points. Black solid lines show arbitrary \OIIIL flux levels. Spaxels with SN $\le$ 3 are masked. Emission lines are corrected for dust attenuation.}
    \label{fig:diagnostic_diagram}
\end{figure}

We computed the following line ratios in all the spaxels with SNR > 5:
\begin{eqnarray}
     \rm O33 = \OIIIL[4363] / \OIIIL \\
     \rm O3\gamma = \OIII[4363] / \Hgamma \\
     \rm R3 = \OIIIL / \Hbeta,   \label{O33_equation}
\end{eqnarray}
\noindent
We adopted the $\rm O3\gamma$ vs $\rm O33$ diagram from \cite{Mazzolari2024} correcting the intensity of the emission lines for dust attenuation (see Sec. \ref{Subsec_bo_dust_extinction}).
 Fig.~\ref{fig:diagnostic_diagram} shows the diagnostic diagram, together with a map of the most likely excitation source in CR7. In this diagram, lying above the demarcation line is a sufficient but not necessary condition for an object to be identified as an AGN, therefore spaxels which are below the demarcation line can still be associated with AGN excitation. We observe that CR7-B and CR7-C lie in the region associated to SF or AGN ionisation. On the other hand, CR7-A shows two regions where the AGN ionisation is likely dominant, with respect to SF ionisation. Nevertheless, we observe that the spaxels associated to AGN ionisation in CR7-A are distributed towards the outskirts of CR7-A and thus have lower SN (SN $\sim$ 3-5) compared to the central region. Additionally, due to the large uncertainties on the line ratios we cannot provide a secure determination of the main source of ionisation in these two regions. Overall, as shown in Fig.~\ref{fig:diagnostic_diagram}, all the points lie preferentially below the demarcation line, thus indicating that the most likely source of ionisation cannot be conclusively stated as either AGN or SF in any of the three components of CR7. 

To trace regions of high ionisation we also computed the R3 line ratio map shown in Fig. \ref{fig:O3_over_Hbeta}. Due to the difference in ionisation potential of \OIIIL and \Hbeta, this ratio is a fairly good indicator of the strength of radiation field, with peaks associated to burst of SF or AGN. Interestingly, Fig. \ref{fig:O3_over_Hbeta} shows a peak of the ratio on CR7-B towards the south, which corresponds to the location where \cite{Sobral2019} detected \HeIIL emission at 2$\sigma$, possibly indicating the presence of an AGN \citep[see also][]{Sobral2015a}. Additionally, as also shown in Fig. \ref{fig:mom_mapso3}, the southern region of CR7-B shows enhanced velocity dispersion in the \textit{broad} component and blue-shifted extended ionised emission which in Sec. \ref{Subsec3_kinematic} we attributed to the hypothetical presence of an outflow (see also Fig.\ref{fig:narrow_broad_detached}). Overall, such region could be ionised by the presence of an AGN, and thus points towards the presence on an AGN-driven outflow, consistent with previous works \citep{Sobral2019}. Similarly to the result inferred from Fig. \ref{fig:diagnostic_diagram} for CR7-A, we observe an enhancement of the line ratio towards the South with peaks of log(R3) $\sim$ 1 at SN $\ge$ 5, suggesting the presence of gas ionised by an AGN. Finally, CR7-C shows low values of R3, consistent with SF ionisation.

\subsection{Low-resolution data - Emission line fitting}\label{Subsec_prism_line_fitting}
\begin{figure}
	\includegraphics[width=\linewidth]{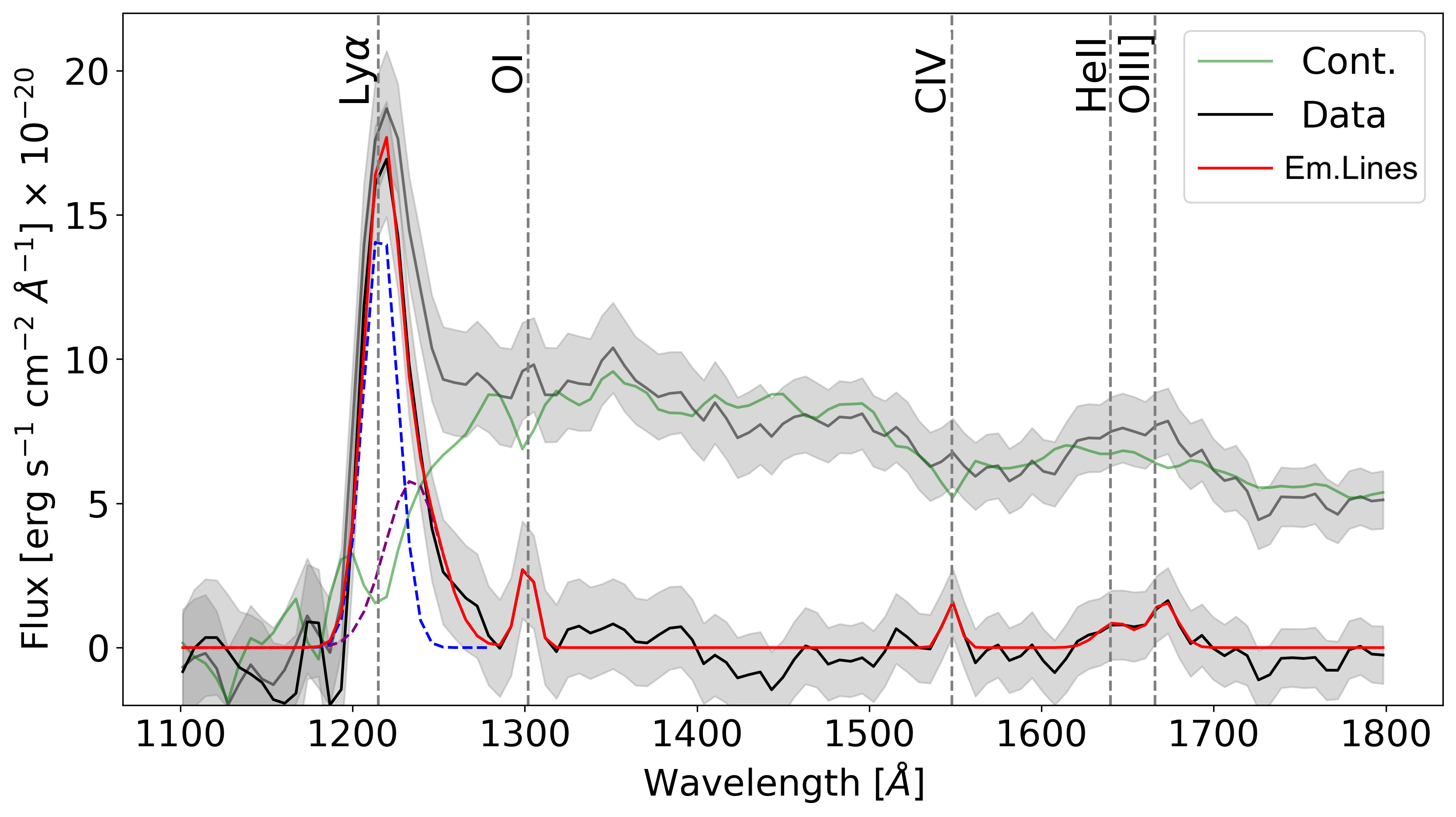}
    \caption{Fit of the integrated R100 spectrum of CR7-A extracted from the aperture shown in Fig. \ref{fig:total_images} and focused on the wavelength range between 1100 - 1800 \AA rest-frame. Data before and after continuum subtraction are in grey and solid black, respectively. Single Gaussians used to fit the \Lyalpha emission line are in dashed blue and purple. Best-fit continuum and emission line models are in green and red, respectively. For details of the spectrum fit see Sec. \ref{Subsec3_continuum_fitting}.}
    \label{fig:Lya_fit_spectrum}
\end{figure}

To analyse the low-resolution data cube we adopted the same methodology described in Sec. \ref{Subsec3_line_fitting}, with a few modifications. In particular, we performed the Voronoi binning on the continuum level between 1100 and 6900 \AA rest-frame, requiring a minimum SN in each spectral channel of 3. Then, to account for the continuum shape of the prism spectrum we used \textsc{pPXF}, simultaneously fitting multiple templates. The best fit is obtained with a linear combination of stellar templates, single Gaussians templates to account for emission line features, and a separate single Gaussian template for the \Lyalpha emission line. To reproduce the observed steep continuum shape (see Fig. \ref{fig:spettri_grism_prism}) we used as input in \textsc{pPXF} the simple stellar population spectra from the stellar-population synthesis tool \textsc{fsps} \citep{Conroy2009, Conroy2010}, covering the total observed spectral range. We used MIST isochrones \citep{Choi2016} and MILES stellar library with fixed solar abundances \citep{SanchezBlazquez2006, Falconbarroso2011}. Then, we included a third order additive polynomial and a sixth order multiplicative polynomial to adjust the continuum shape of the stellar templates to the observed spectra. These polynomials are necessary to fully capture the continuum shape \citep[e.g.,][]{bezanson2018}, but we do not physically interpret the resulting continuum fits. Then, similarly to the routine described in Sec. \ref{Subsec3_line_fitting}, we subtracted the best-fit continuum from the observed spectrum in each spaxel and proceed to fit the emission lines. Due to the low spectral resolution of the prism data we only adopted a single Gaussian component to reproduce the emission line features. Finally, due to the asymmetric profile of \Lyalpha we performed the fit of this emission line with a combination of two Gaussians. Fig. \ref{fig:Lya_fit_spectrum} shows the integrated spectrum extracted from CR7-A between 1100 - 1800 \AA rest-frame, with the best fit continuum and multi-Gaussian model. The continuum subtraction highlighted the presence of multiple high-ionisation emission lines, most of them detected at the level of only 2 $\sigma$. We measured a total \Lyalpha integrated flux of 6.4 $\pm$ 0.8 $\times$ 10$^{-17}$ erg s$^{-1}$ cm$^{-2}$, which is smaller of a factor of $\sim$ 1.6 compared to MUSE observations \citep{Matthee2020}. We remark that the flux of \permittedEL[O][i][\textlambda][1302] and \CIVall is degenerate with the underlying stellar absorption at the same wavelength. In this work we found no evidence for \HeIIL with high equivalent width in any of the satellites. In particular, we estimated an upper limit on the non detection in CR7-A of $\leq$ 0.3 $\times$ 10$^{-18}$ erg s$^{-1}$ cm$^{-2}$, which is consistent with the upper limit of  $\leq$ 0.27 $\times$ 10$^{-18}$ erg s$^{-1}$ cm$^{-2}$ provided by \citet{Sobral2019}. Nevertheless, \citet{Sobral2019} also reported a 2$\sigma$ detection of \HeIIL over CR7-C with an integrated flux of 0.11 $\pm$0.5 $\times$ 10$^{-18}$ erg s$^{-1}$ cm$^{-2}$, which is in contrast with our non detection.

%

\subsection{Low-resolution data - Continuum fitting}\label{Subsec3_continuum_fitting}
In this section we describe the methodology we adopted to fit the continuum emission in CR7, to finally provide spatially resolved stellar population properties. We follow the methodology developed in \citet{Marconcini2024}. We performed the continuum fitting of the low-resolution data cube in three different manners, i.e. spaxel by spaxel, by fitting the total integrated spectrum and fitting the spectra of CR7-A, CR7-B, and CR7-C, separately. To fit the spectrum we used \textsc{prospector} \citep[][]{Johnson2021}, a Bayesian SED modelling framework built around the stellar-population synthesis tool \textsc{fsps} \citep{Conroy2009, Conroy2010}.
We performed a SED modelling of the observed spectrum from 0.6 to 5.27~$\mu$m following the procedure described in \citet{Tacchella2023} and \cite{perez_gonzales2023}. In order to fit the spectrum with \textsc{prospector} we assumed a 2-d Gaussian kernel, with different full-width half-maximum values along and across slices, as a function of wavelength and performed a spatial smoothing, in order to obtain approximately the same PSF at all wavelengths \citep{deugenio2023}. 

\begin{figure}
    \includegraphics[width=\linewidth]{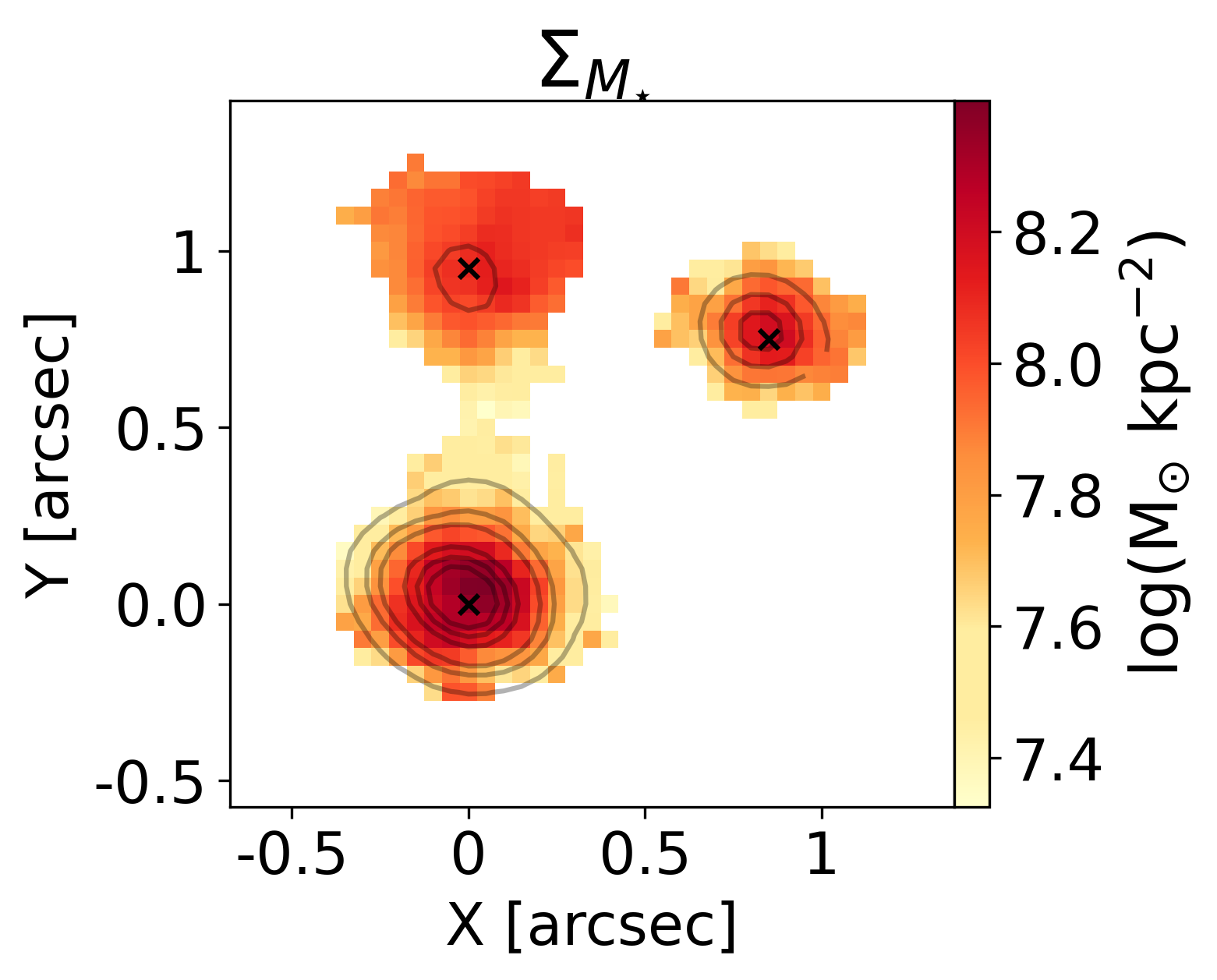}
    \caption{Map of the stellar mass surface density derived with \textsc{prospector}. The black crosses mark the positions of the three UV bright components of CR7. Black contours are arbitrary \OIIIL flux levels.}
    \label{fig:stellar_mass_prospector}
\end{figure}

For the modelling, we configured \textsc{fsps} to use the MILES stellar atmospheres \citep{Falconbarroso2011} and MIST isochrones \citep{Choi2016}. The nebular emission is modelled using pre-computed grids from \textsc{cloudy} \citep{Ferland1998}, as described in \citet{Byler2017}; this approach takes into account possible stellar absorption at the wavelength of emission lines \citep[see e.g.][]{perez_gonzales2003, perez_gonzales2008}. Finally, we accounted for dust attenuation using a flexible dust attenuation law, consisting of a modified Calzetti law \citep[][]{Calzetti2000} with a variable power-law index and UV-bump strength \citep{Noll2009, Kriek2013}. Stars younger than 10~Myr are further attenuated by an extra dust screen, parametrised as a simple power law \citep{Charlot2000}. The star-formation history (SFH) uses 9 fixed time bins between z = 6.60425 and z = 20; the first three bins are at 10~Myr, 30~Myr, and 100~Myr, the remaining 6 bins are logarithmically spaced in time. We use a continuity prior to relate the log ratio of the SFRs between adjacent time bins \citep{Leja2019}.
The model free parameters and their prior probability distributions for CR7-A, CR7-B, and CR7-C are listed in Table~\ref{t.prospector}. Fig.~\ref{fig:stellar_mass_prospector} shows the spatially resolved stellar mass surface density ($\Sigma_{\rm M_{\star}}$) estimated with this method, with a clear peak of stellar mass formed in CR7-A.

\begin{figure}
    \includegraphics[width=\linewidth]{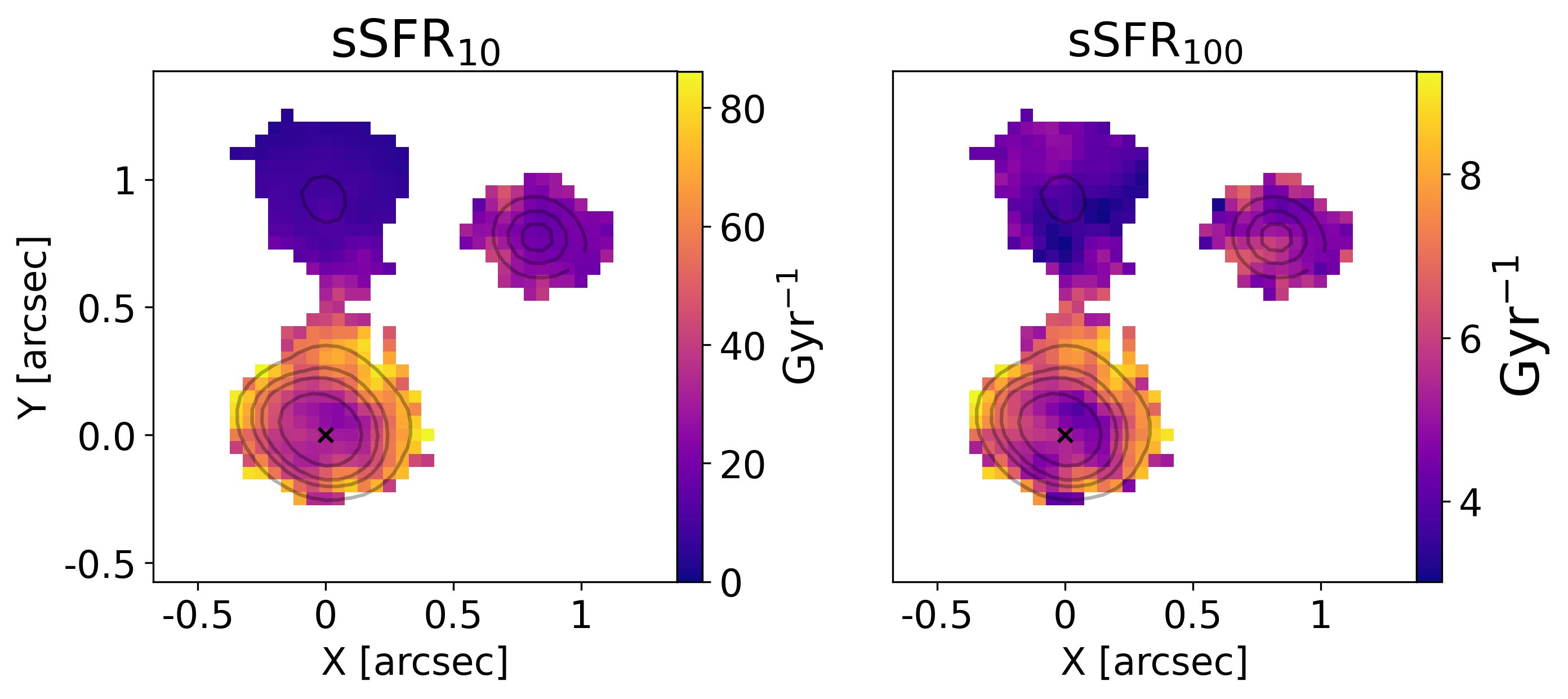}
    \caption{Specific Star formation rate (sSFR) within the last 10 Myr (Left) and 100 Myr (Right). Black contours are arbitrary \OIIIL flux levels, as in Fig. ~\ref{fig:sfr_10_100_Ha}.}
    \label{fig:ssfr_prospector}
\end{figure}
\subsubsection{Stellar population properties}\label{Subsub_sec_stellar_pop_properties}
From the integrated spectrum of the entire CR7 system, we found a total stellar mass budget of $\log(\mstar/\msun)_{\rm CR7}=9.29^{+0.17}_{-0.16}$  which is consistent with the value we obtain from modeling the spectrum in each spaxel and then adding up the resulting stellar masses (which gives $\log(\mstar/\msun)=9.44$). From the integrated spectrum of each component, instead, we found $\log(\mstar/\msun)_{\rm CR7-A}=9.33^{+0.05}_{-0.07}$, $\log(\mstar/\msun)_{\rm CR7-B}=8.52^{+0.16}_{-0.15}$, and $\log(\mstar/\msun)_{\rm CR7-C}=8.29^{+0.31}_{-0.11}$ (see Tab. \ref{t.prospector}). The estimated stellar mass of CR7-A is consistent with a previous SED fitting estimate of $\log(\mstar/\msun)_{\rm CR7_A}=9.2^{+0.3}_{-0.1}$ based on broad-band photometry \citep{Bowler2017}.
\begin{figure*}
    \includegraphics[width=\linewidth]{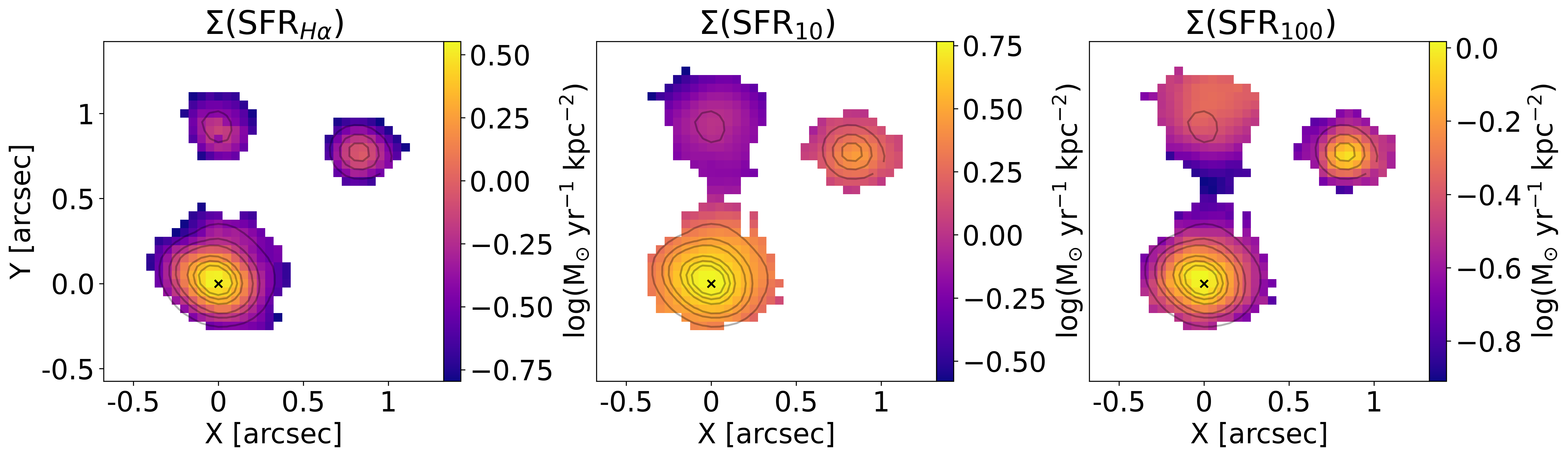}
    \caption{CR7 properties derived from emission line and SED fitting of the low-resolution data cube. From left to right the surface SFR density derived from \Halpha spatially resolved emission (see Eq. \ref{eq.sfr_Ha}), the SFR within the last 10 Myr and 100 Myr derived with \textit{prospector} (see Sec. \ref{Subsec3_continuum_fitting}). Black contours are arbitrary \OIIIL flux levels.}
    \label{fig:sfr_10_100_Ha}
\end{figure*}
From the SED fitting of the integrated spectrum we derived a SFR$_{10} = 55^{+1}_{-1}$ \msun yr$^{-1}$ and SFR$_{100} = 9.1^{+1.2}_{-1.2}$ \msun yr$^{-1}$, integrating the SF history (SFH) of the last 10 and 100 Myr, respectively. Combining the result for the total stellar mass formed and the SFR, we computed the sSFR within the last 10 and 100 Myr. The resulting spatially resolved maps are shown in Fig. \ref{fig:ssfr_prospector}. From the integrated spectrum, instead, we derived a specific SFR of sSFR$_{10}$ = 29$^{+13}_{-10}$ Gyr$^{-1}$ and sSFR$_{100}$ = 4.91 $^{+1.19}_{-1.15}$ Gyr$^{-1}$. The sSFR maps show extremely different values among the components of CR7, with higher sSFR surrounding CR7-A, and lower values in the satellites. This pattern is evident both in the sSFR$_{10}$ and sSFR$_{100}$ maps. Nonetheless, we observed higher sSFR over CR7-C integrating over the last 100 Myr, compared to the last 10 Myr only, indicating that most mass of CR7-C is formed more than 10 Myr ago. 

As a comparison with the SFR derived via SED fitting, we measured the \Halpha flux from the low-resolution data to provide a direct estimate of the SFR in CR7. We assumed a Case B recombination scenario \citep{Osterbrock2006} and based on the measured \Halpha flux of 4.4 $\pm$ 0.2 $\times$ 10$^{-17}$ erg s$^{-1}$ cm$^{-2}$, we computed the SFR following the relation proposed by \citep{Murphy2011}, as:
\begin{equation}
    \rm SFR_{\Halpha} = 5.4 \times 10^{-42} \  \left( \frac{L_{\Halpha}}{erg s^{-1}}\right) \ \msun \mathrm{yr}^{-1}
\end{equation}\label{eq.sfr_Ha}\noindent
Then, we derived the spatially resolved $SFR_{\Halpha}$ density map ($\Sigma (SFR_{\Halpha})$) as shown on the left panel of Fig. \ref{fig:sfr_10_100_Ha}. On average, we found a log(SFR$_{\Halpha}$/\msun yr$^{-1}$) =  1.5 $\pm$ 0.4. Similarly as for the \Halpha, we traced the SFR via \Hbeta emission and found log(SFR$_{\Hbeta}$/\msun yr$^{-1}$) =  1.4 $\pm$ 0.5 , which is consistent both with SFR$_{\Halpha}$ and the SFR derived via SED fitting (see Tab. \ref{t.prospector}).

\begin{figure*}
	\includegraphics[width=\linewidth]{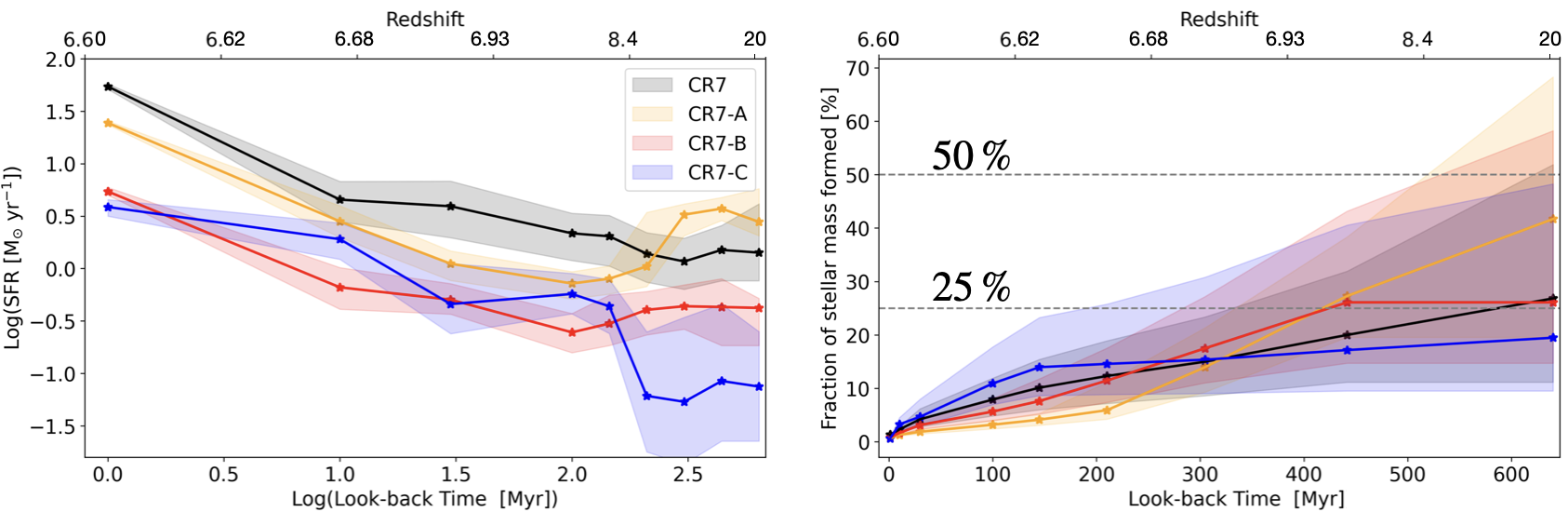}
    \caption{Star formation history (left) and fraction of stellar mass fraction formed per bin of time as a function of the look-back time (right) in CR7. The properties of the total spectrum of CR7, and for CR7-A, CR7-B, and CR7-C are shown in black, orange, red, and blue, respectively. Errorbars represent the 84\textsuperscript{th} and 16\textsuperscript{th} percentiles. Horizontal black dashed lines in the right panel mark the level when 50 \% and 25 \% of the stellar mass was formed in CR7.}
    \label{fig:sfh}
\end{figure*}

Fig. \ref{fig:sfh} shows the SFH and the fraction of stellar mass formed as a function of look-back time in CR7. The SFH of the integrated spectrum of CR7 shows that the SFR is slightly decreasing toward higher redshift. Interestingly, we noticed that the SFH of CR7-A and CR7-C anti-correlate at z $\ge$ 8, potentially indicating an interaction-driven star formation. Indeed, between look-back times of $\sim$ 200 and 150 Myr ago, the SFRs of CR7-A and CR7-C decrease and increase, respectively, following a mirrored trend. Such mirrored SF is consistent with predictions from simulations, which suggest that strong interactions among merging companions can lead to out-of-phase SF due to alternating gas inflows and feedback suppression effects \citep{Hani2020, Renaud2022}. Then, from $\sim$100 up to 10 Myr ago, their SFRs are consistent within the uncertainties. We interpret such a mirrored trend of the SFH of CR7-A and CR7-C in Fig. \ref{fig:sfh} as due to gas transfer among these two components. In particular, as already discussed in Sec. \ref{Subsec3_kinematic}, we clearly observe a strong interaction between CR7-A and CR7-C due to elongated structures of ionised gas connecting them. Therefore, a scenario consistent with the observed SFH (Fig. \ref{fig:sfh}) and with previous findings would be that gas is accreted from CR7-C onto CR7-A, boosting the SFR at early times. Conversely, due to the loss of gas the SFR in CR7-C is low. Such a scenario suggests a strong interaction between CR7-A and CR7-C at look-back times $\geq$ 200 Myr, followed by a less extreme interaction at recent times, as supported by the ionised gas kinematics and elongated morphology (see Fig. \ref{fig:mom_mapso3}). The SFH of CR7-B instead has an approximately constant SFR throughout all its history. The fraction of the stellar mass formed shows an average identical trend for all the components of CR7. In particular, CR7-A already formed $\geq$ 80 \% of its total stellar mass more than 300 Myr ago, followed by a drastic decrease of star formation within the last $\sim$ 200 Myr. Overall, the total fraction of stellar mass formed shows a slightly increasing trend, up to z $\sim$ 9.3, where $\sim$ 25 \% of the total stellar mass of CR7 was already formed. 

\subsubsection{The escape fraction of ionising photons}\label{Sub_sub_sec_escape_photons}
\begin{figure}
    \includegraphics[width=\linewidth]{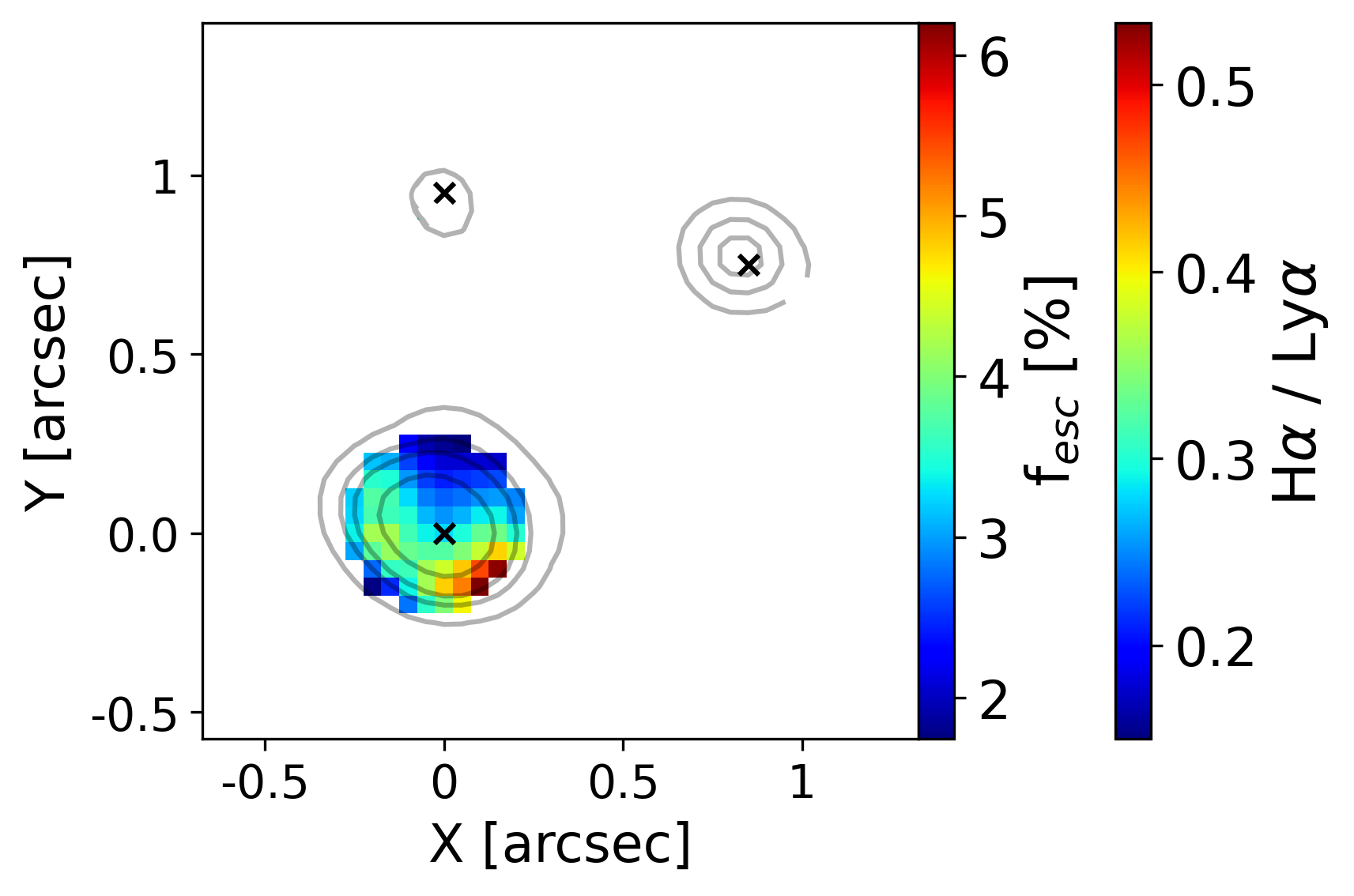}
    \caption{Resolved map of ionizing photon production. Different colorbars indicate the \Halpha/\Lyalpha emission line ratio in logarithmic scale and the escape fraction of LyC ionizing photons as derived from the \Halpha/\Lyalpha ratio, with the method outlined in Sec. \ref{Sub_sub_sec_escape_photons}.}
    \label{fig:escape_frac_and_lya_ha_ratio}
\end{figure}

Ionising LyC photons can escape the galaxy through cavities of the ISM and ionise the neutral gas or be absorbed by dust. There is a strong connection between the escape fraction of such ionizing photons and the \Lyalpha flux. Here we estimate the fraction of escaping LyC photons in CR7 following the methodology described in \citet{Matthee2017_escape_photons}. Adopting the dust fraction ($f_\mathrm{dust}$) derived via SED fitting (see Sec. \ref{Subsec3_continuum_fitting}) and assuming case B recombination we can write the fraction of escaping LyC photons as:
\begin{equation}
    f_\mathrm{esc} = \frac{1 - f_\mathrm{dust}}{\left(1 + \alpha \frac{L_{\Halpha}}{L_{\Lyalpha}}\right)}
\end{equation}\label{eq.escape_fraction}
\noindent
where $\alpha$ is a parameter that depends on the absorption of ionising photons, the average energy of ionising photons, and the \Halpha emission coefficient \citep[for details see][]{Matthee2017_escape_photons}. $L_{\Halpha}$ and $L_{\Lyalpha}$ are the \Halpha and \Lyalpha luminosities, respectively. In this work from low-resolution data fitting we estimated $L_{\Halpha}$ = 2.31 $\pm$ 0.01 $\times 10^{43}$ erg s$^{-1}$ and $L_{\Lyalpha}$ = 3.34 $\pm$ 0.04 $\times 10^{43}$ erg s$^{-1}$, which is slightly lower than the the $L_{\Lyalpha}$ estimated by previous MUSE observations \citep{Matthee2020}. The ratio of \Halpha/\Lyalpha and the derived LyC escape fraction in each spaxel are shown in Fig. \ref{fig:escape_frac_and_lya_ha_ratio}. In contrast to previous works \citep{Matthee2020}, we detect \Lyalpha emission only surrounding CR7-A. We observe a north-south gradient of $f_\mathrm{esc}$, with values up to 6 \% towards the South in CR7-A, which has to be ascribed to a gradient in the ratio of \Halpha/\Lyalpha. On average, we estimated a fraction of escaping LyC photons of $f_\mathrm{esc}$ = 3.4 $\pm$ 0.9 \%. Fig. \ref{fig:escape_frac_and_lya_ha_ratio} shows a remarkable maximum of the escape of LyC photons towards the southern part of CR7-A, which is interestingly co-spatial to the region of enhanced \OIIIL/\Hbeta ratio (Fig. \ref{fig:O3_over_Hbeta}). We recall from Sec. \ref{Narrow_broad_section} that the \textit{broad} emission is elongated in the NE-SW direction with enhanced velocity dispersion directed North-South and was plausibly associated with an outflow, driven either by SF or an AGN, pushing the ambient gas towards the outskirts of CR7-A. Such spatial agreement could indicate a potential causal relationship between the outflow activity and the enhanced ionizing photon escape. The observed correlation can be understood in the context of feedback mechanisms, where outflows are known to impact the ISM by injecting energy and momentum, creating low-density cavities within the ISM, and thus reducing the optical depth along these paths. This decrease in gas, and hence column density might ease the escape of ionizing photons within outflow regions. Such a mechanism is suggested both in theoretical models and simulations, where AGN-driven winds are expected not only to clear the gas content of the ISM but to propagate along the path of least resistance, allowing ionizing radiation to escape more efficiently into the IGM \citep{Murray2011, Hopkins2012}. Additionally, the kinetic energy injected by the outflow into the ISM can evacuate the outflow cavity, further diminishing the covering fraction of dense material along the line of sight. As a consequence, the average column density of the gas decreases, thereby increasing the transparency of the ISM to ionizing photons. Such an outcome is consistent with the scenario where AGN activity not only quenches star formation by heating and expelling gas but also contributes to the reionization of the IGM by increasing the escape of ionizing radiation \citep{Costa2014, Gabor2014}.


\begin{table*}
    \begin{center}
    \caption{Summary of the parameters, prior probabilities and posterior probabilities of the fiducial SED fitting \texttt{prospector} model (see also Fig.~\ref{fig:stellar_mass_prospector}).
    }\label{t.prospector}
    \setlength{\tabcolsep}{3pt}
    \begin{tabular}{llcp{4.5cm}llcllc}
  \hline
   & Parameter & Free & Description & Prior & CR7 & CR7-A &  CR7-B & CR7-C \\
   & (1)       & (2)  & (3)         & (4)   & (5) & (6) & (7)& (8)  \\
   \hline
   \multirow{8}{*}{\rotatebox[origin=c]{90}{Free parameters}}
   & $\log \mstar[\msun]$ & Y & total stellar mass formed & $\mathcal{U}(7, 13)$ & $9.29^{+0.17}_{-0.16}$ & $9.33^{+0.05}_{-0.07}$ & $8.52^{+0.16}_{0.15}$ & $8.29^{+0.31}_{-0.11}$  \\
   & $\log Z [\zsun]$ & Y & stellar metallicity & $\mathcal{U}(-2, 0.19)$ & $-1.03^{+0.11}_{-0.11}$ & $-1.37^{+0.08}_{-0.07}$ & $-1.22^{0.12}_{-0.12}$ & $-0.42^{+0.26}_{-0.49}$   \\
   & $\log \mathrm{SFR}$ ratios & Y & ratio of the $\log \mathrm{SFR}$ between adjacent SFH bins & $\mathcal{T}(0, 0.3, 2)$ & --- & --- & --- & ---\\
   & $\tau_V$ & Y & optical depth of the diffuse dust & $\mathcal{G}(0.3,1;0,2)$ & $0.07^{+0.0 1}_{-0.01}$ & $0.18^{+0.01}_{-0.01}$ & $0.13^{+0.05}_{-0.03}$ & $0.07^{+0.04}_{-0.03}$  \\
   & $\mu$ & Y & ratio of the optical depth of the birth clouds to $\tau_V$ & $\mathcal{U}(-1.0,0.4)$ & $1.12^{+0.07}_{-0.07}$ & $0.43^{+0.10}_{-0.12}$ & $0.72^{0.17}_{-0.23}$  & $0.79^{4}_{-0.19}$ \\
   & $\sigma_\mathrm{gas} \; [\kms]$ & Y & intrinsic velocity dispersion of the star-forming gas$^\ddag$ & $\mathcal{U}(0,300)$ &  $165^{+40}_{-50}$ & $124^{+50}_{-18}$ & $140^{+20}_{-30}$ & $108^{+50}_{-27}$ \\
   & $\log Z_\mathrm{gas} [\zsun]$ & Y & metallicity of the star-forming gas & $\mathcal{U}(-2, 0.19)$ & $-0.59^{+0.01}_{-0.01}$ & $-0.59^{+0.01}_{-0.02}$ & $-0.64^{+0.05}_{-0.05}$ & $-0.40^{+0.06}_{-0.12}$  \\
   & $\log U$ & Y & ionisation parameter of the star-forming gas & $\mathcal{U}(-4, -1)$ & $-1.18^{+0.06}_{-0.07}$ & $-1.01^{+0.01}_{-0.02}$ & $-1.41^{+0.13}_{-0.11}$ & $-1.54^{+0.25}_{-0.18}$  \\
   \hline
   \multirow{2}{*}{\rotatebox[origin=c]{90}{Other}}
   & $\log SFR_{10} [\msun \, \peryr]$ & N & star-formation rate averaged over the last 10~Myr & --- & $1.74^{+0.02}_{-0.03}$ & $1.39^{+0.02}_{-0.02}$ & $0.73^{+0.04}_{-0.03}$ & $0.58^{+0.07}_{-0.09}$  \\
   & $\log SFR_{100} [\msun \, \peryr]$ & N & star-formation rate averaged over the last 100~Myr & --- & $0.96^{+0.09}_{-0.08}$ & $0.58^{+0.05}_{-0.05}$ & $0.01^{+0.06}_{-0.06}$ & $0.06^{+0.16}_{-0.10}$   \\
  \hline
  \end{tabular}
  \end{center}
(1) Parameter name and units (where applicable). (2) Only parameters marked with `Y' are optimised by \texttt{prospector}; parameters marked with `N' are either tied to other parameters (see Column~4), or are calculated after the fit from the posterior distribution (in this case, Column~4 is empty). (3) Parameter description. For the dust attenuation parameters $n$, $\tau_V$ and $\mu$ see \citet[][their eq.s~4 and~5]{Tacchella2022}. (4) Parameter prior probability distribution; $\mathcal{N}(\mu, \sigma)$ is the normal distribution with mean $\mu$ and dispersion $\sigma$; $\mathcal{U}(a, b)$ is the uniform distribution between $a$ and $b$; $\mathcal{T}(\mu, \sigma, \nu)$ is the Student's $t$ distribution with mean $\mu$, dispersion $\sigma$ and $\nu$ degrees of freedom; $\mathcal{G}(\mu, \sigma, a, b)$ is the normal distribution with mean $\mu$ and dispersion $\sigma$, truncated between $a$ and $b$.
(5), (6), (7), and (8) are the posterior median and 16\textsuperscript{th}--84\textsuperscript{th} percentile range of the marginalised posterior distribution for CR7-A, CR7-B, CR7-C, and total CR7, respectively; for some nuisance parameters we do not present the posterior statistics (e.g., log SFR ratios). $^\ddag$ The velocity dispersion of the emission lines is a nuisance parameter, due to the low spectral resolution of the prism data ($\sim 500$~\kms at the wavelength of \OIIIL); indeed, the posterior probability distributions for CR7 is consistent with 0~\kms.
\end{table*}



\subsection{Gas metallicity}\label{Subsec3_metallicity}
Investigating the gas metallicity in a merging system such as CR7 is crucial to comprehend how the distribution of metals results from the gas exchange among the components, which leads to bursts of SF and thus further metal enrichment. Unfortunately, we do not detect the \SIIall doublet neither in the high- or low-resolution data and the spectral coverage of the G395H data does not include the \OIIall doublet. As a consequence, we cannot estimate the electron density from observed emission lines ratios. Therefore, we assumed an average electron density of 400 cm$^{-3}$ based on estimates from sources at similar redshift and on the observed electron density redshift evolution \citep[][]{Isobe2023, Abdurrouf2024, Marconcini2024} and then used $Pyneb$ to derive the collision strengths of oxygen ions with the default atomic dataset \textit{PYNEB\_21\_01}. Then, we computed the electron temperature (T$_e$) from the \OIIIL/\OIIIL[4363] line ratio. We computed the total oxygen abundance in each spaxel where we could estimate the T$_e$, assuming that the oxygen abundance is representative of the total gas metallicity \citep[for a detailed discussion on the underlying assumptions see][]{Maiolino2019}.

\begin{figure*}
    \includegraphics[width=\linewidth]{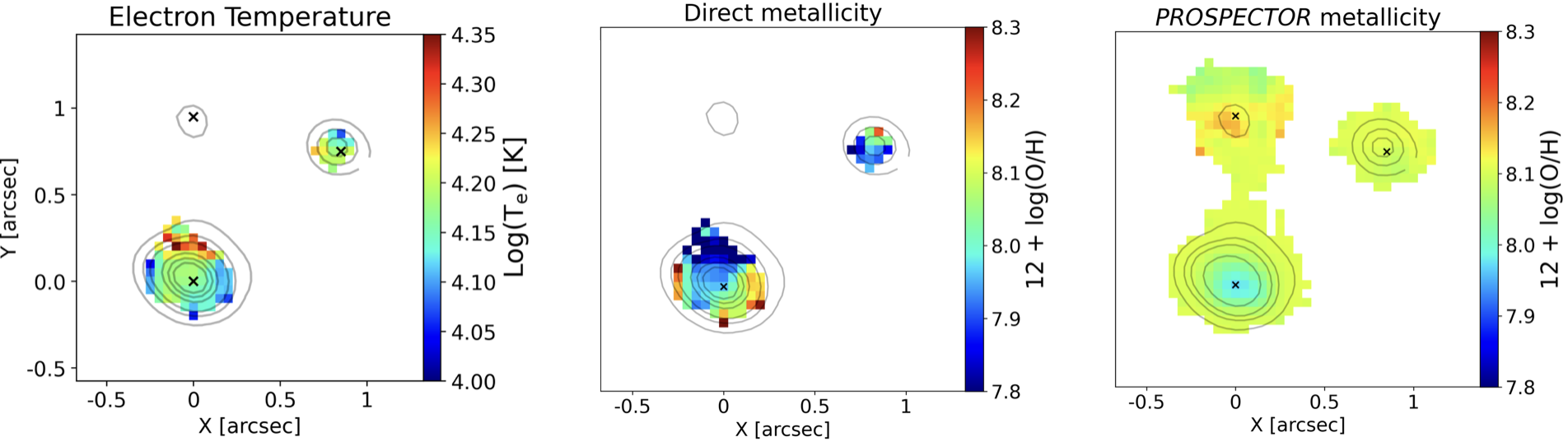}
    \caption{ISM physical properties of CR7. From left to right: Electron temperature, gas-phase metallicity derived with the direct method (T$_\mathrm{e}$ method) and gas-phase metallicity derived with \textsc{prospector} performing SED fitting (see Sec. \ref{Subsec3_line_ratios} and Sec. \ref{Subsec3_continuum_fitting} for details). Spaxels at SN $\le$ 3 for the \OIIIL[4363] are masked. Black contours are arbitrary \OIIIL flux levels. The black crosses mark the positions of the three components.}
    \label{fig:metalliticities}
\end{figure*}

Fig. \ref{fig:metalliticities} shows spatially resolved maps of the gas electron temperature (left) and the gas-phase metallicity derived with the direct T$_e$ method (middle) and with \texttt{prospector} (right; the latter uses pre-computed photoionization grids, see Sec. \ref{Subsec3_continuum_fitting}). The average electron temperature estimated from the integrated spectra of CR7-A and CR7-B are 1.6 and 0.9 $\times$ 10$^{4}$K, respectively. As shown on the right panel in Fig. \ref{fig:metalliticities}, CR7-A has the lowest metal content in the entire system, with an average metallicity of 12 + log(O/H) = 8.0, i.e Z = 0.2 \zsun, consistently with previous works \citep{Matthee2017b_alma, Dors2018}. On the other hand, CR7-B and CR7-C have average metallicities from direct T$_e$ method of 8.08 and 8.26, respectively (see also Tab. \ref{t.prospector}). Due to the low SN of the \OIIIL[4363] emission line in CR7-C we could not resolve its emission from the high-resolution data cube. Therefore, we computed the \OIIIL[4363] integrated line profile and estimated an average direct-T$_e$-method gas phase metallicity of 12 + log(O/H) = 8.0 $\pm$ 0.5. 
In contrast with the gas-phase metallicity distribution derived with the direct-T$_e$ method (middle panel), the \textsc{prospector} fit (right panel) in Fig. \ref{fig:metalliticities} does not show any metallicity gradient in CR7-A. We ascribe such a different pattern to the different methods used to derive the gas-phase metallicity. In the central panel the metallicity   is derived from ``direct'' $T_e$ method. In the right panel the metallicity is derived by \textsc{prospector}, which relies on \textsc{cloudy} models to derive the nebular emission \citep{Ferland1998}. In the latter case it is important to note that, in order to perform the fit with \textsc{prospector}, we convolved the low-resolution data cube with the wavelength-dependent PSF (see \citet{deugenio2023} and Sec.\ref{Subsec3_continuum_fitting}), which can possibly result in a loss of spatial informations. Such differences regarding the method to derive the gas metallicity are plausibly causing the observed difference.

Nevertheless, we observe that on average the gas-phase metallicity estimated with the direct-T$_e$ method and the Prospector fitting analysis provide consistent values for each satellites in the system. Moreover, from the total integrated spectrum of CR7 we found an average metallicity of Z = 0.2 $\pm$ 0.08 \zsun and Z = 0.26 $\pm$ 0.01 \zsun from direct T$_e$ method and SED fitting, respectively. Overall, our findings for both the direct T$_e$ and the gas phase metallicity estimated with \textsc{prospector} are consistent with the analysis of \cite{Matthee2017b_alma} and \citet{Dors2018} that measured an integrated metallicity of Z = 0.05 - 0.2 \zsun based on multiple methodologies employing \CIIL luminosity and UV lines, SFR and photoionization modelling, respectively.

We observed a metallicity gradient across CR7-A, showing higher values towards the South-West edge, opposite to the electron temperature trend (see Fig. \ref{fig:metalliticities}). This region of enhanced metal content in CR7-A is co-spatial with the hypothetical high-excitation region (see Fig. \ref{fig:diagnostic_diagram}) and the region of high \OIIIL/\Hbeta ratio. As discussed in Sec. \ref{Subsec3_line_ratios}, the presence of an outflow emerging from CR7-A is consistent with the observed features. Interestingly, both models and simulations predict higher values of metal content co-spatial to AGN-driven outflows, under the scenario of a wind pushing the inner metal-rich gas towards the outskirts of the galaxy, thus causing the enhanced metal content within the outflow \citep{Moll2007, Germain2009, Shen2010, Taylor2015, Choi2020}. As discussed in Sec. \ref{Subsec3_line_ratios}, in this work we cannot conclusively state or rule-out the presence of an AGN in the centre of CR7-A, which in case would make our estimates of the gas-phase metallicity unreliable.

Interestingly, the right panel in Fig. \ref{fig:metalliticities}, showing the metallicity measured with \textsc{prospector}, reports a higher gas-phase metallicity in CR7-C, with an average value of 12 + log(O/H) = 8.12 $\pm$ 0.02. As we will discuss in the following section, our findings show unambiguous evidence of vigorous interaction between CR7-A and CR7-C, with gas being transferred towards the most massive galaxy, i.e., CR7-A. In such a scenario, preferential accretion of gas onto CR7-A instead of CR7-C would lower the metal content of CR7-A, by diluting the average gas-phase metallicity, as observed. On the other hand, the metallicity in CR7-C should increase due to weak cosmic gas accretion, thus the ISM is enriched in metals as a by-product of stellar evolution that is not counteracted by dilution.

\section{Discussion}\label{Sec4_discussion}
In this paper we presented the spatially resolved and integrated properties of CR7, leveraging the high spectral resolution of the G395H grating and the broad spectral coverage of the low-resolution prism from NIRSpec/IFS data. Here we summarise and discuss the interaction among the various components in CR7 as a result of our analysis, comparing our findings with previous works.

\subsection{Multiple satellites in CR7}\label{subsec4_interacting_companions}
As outlined in Sec. \ref{Sec3_analysis_line_fitting_line_ratios}, multiple lines of evidence, including the ionised gas kinematics, gas-phase metallicity and SFH via continuum fitting, are consistent with a scenario of interaction between the UV-bright components CR7-B, and CR7-C and the most massive galaxy in CR7, i.e. CR7-A, suggesting they all form a merging system. Specifically, as shown in Figs. \ref{fig:rgb_image}, \ref{fig:mom_mapso3} and in agreement with the findings of \citet{Matthee2017b_alma}, we detected multiple components in CR7 that exhibit different projected systemic velocities. To summarise, CR7-C, CR7-C1, CR7-C2, and CR7-A1 display narrow emission lines ($\sigma$ $\leq$130 km s$^{-1}$) , consistent with the interpretation of these clumps as detached satellites falling into the gravitational centre of CR7. CR7-B shows a narrow emission line profile with a blue-shifted wing extending southward, possibly indicative of gas inflow or outflow. The presence of multiple sub-structures, which are likely interacting with the three UV-bright satellites in CR7, aligns well with theoretical models and simulations of structure formation via clump aggregation in the early Universe \citep{Dekel2009_clumps, Mandelker2014, Oklopvic2017}. Moreover, there is growing observational evidence of satellite aggregation towards central galaxies at high redshift, which significantly contributes to the total stellar mass of the main galaxy \citep[][see also Cresci et al. in prep for the Himiko galaxy]{Vanzella2017, Carniani2017, Carniani2018, Carniani2018_cr7, Zanella2019, Duan2024}.
Based on dynamical mass estimates and stellar mass content (see Sec. \ref{Subsec3_kinematic} and Tab \ref{t.prospector}), CR7-A is likely to be the central and most massive galaxy within the system, as previously concluded by several studies \citep{Sobral2015a, Bowler2017, Matthee2017b_alma, Matthee2020}. Moreover, the SFH and the fraction of stellar mass formed as a function of the look-back time (see Fig. \ref{fig:sfh}) suggest that CR7-C and CR7-A began influencing their mutual evolution as early as 600 Myr ago, probably through gas inflow from CR7-C towards CR7-A. By analysing the SFH of the individual satellites, we propose that such a merger caused the transfer of a substantial amount of gas mass until approximately 150 Myr ago, when the SFR of CR7-C and CR7-A began to converge (see Fig. \ref{fig:sfr_10_100_Ha}).

\subsection{Resolved FMR}\label{Subsec_resolved_FMR}
Many works bring evidence of a three-dimensional relationship between M$_\star$-Z$_\mathrm{gas}$-SFR \citep{Tremonti2004, Mannucci2010, Maiolino2019, Cresci2019}, called `Fundamental Metallicity Relation' (FMR) due to its independence from redshift at least up to z$\sim$3 \citep{Cresci2019}. In a recent work, \citet{Baker2023} provided evidence for a spatially resolved version of this relationship, highlighting that local metallicity is primarily influenced by $\Sigma_{M_{\star}}$ and present an inverse relationship with $\Sigma_{SFR}$.

\begin{figure}
    \includegraphics[width=\linewidth]{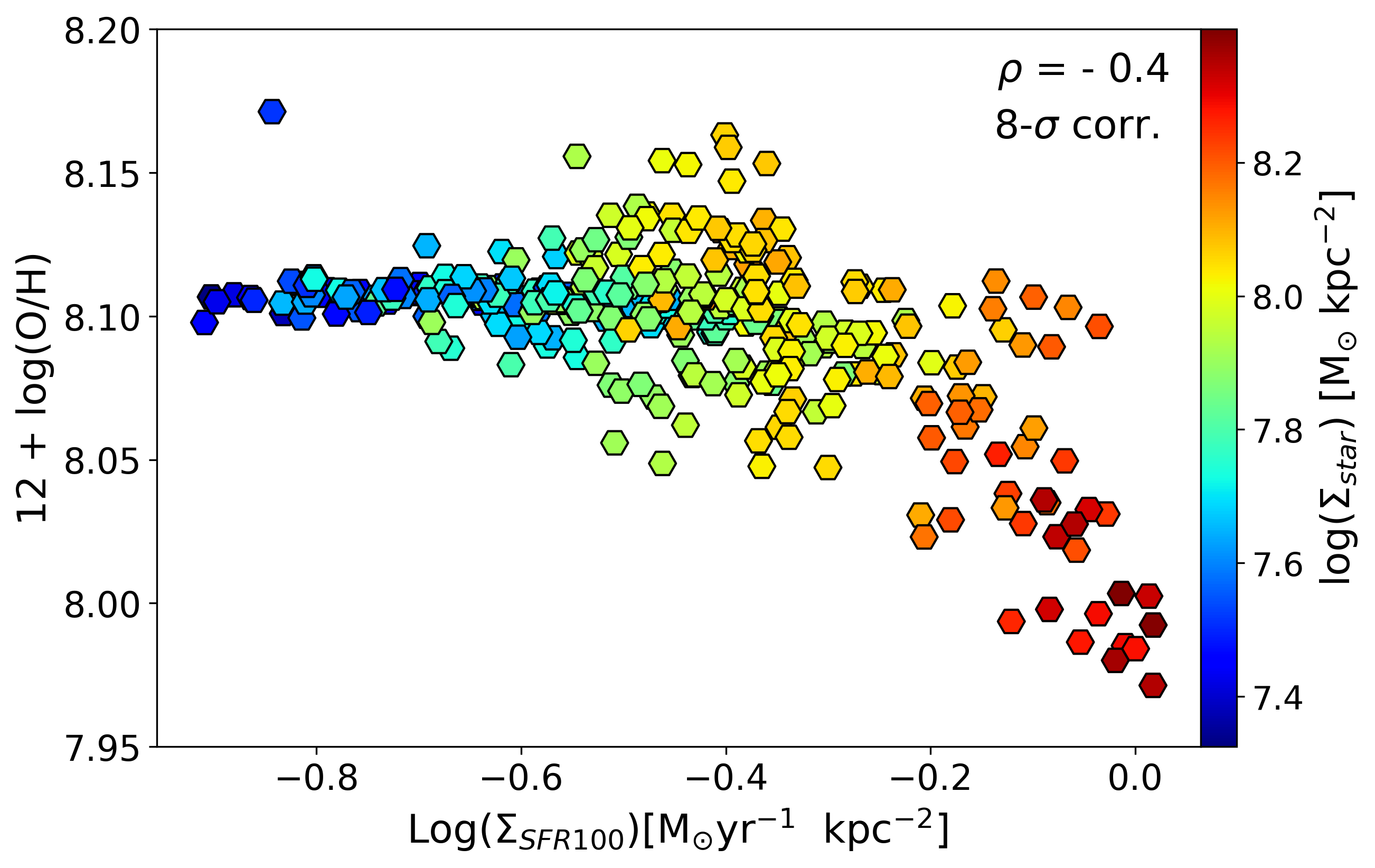}
    \caption{Distribution of the spatially resolved gas-phase metallicity and SFR surface density colour-coded with the surface density of the total stellar mass formed in CR7. All properties are derived with \textsc{prospector} (see Sec. \ref{Subsec3_continuum_fitting} and Figs. \ref{fig:metalliticities}, \ref{fig:stellar_mass_prospector}).}
    \label{fig:anticorrelation_SFR_M}
\end{figure}

In this work, we utilized the resolved maps of SFR$_{100}$ density and gas-phase metallicity as provided by \textsc{prospector} to investigate such observed anti-correlation in CR7 at z$\geq$ 6.6.  Fig.~\ref{fig:anticorrelation_SFR_M} shows the observed anti-correlation between the gas-phase metallicity and SFR$_{100}$ in CR7. As shown in Fig.~\ref{fig:anticorrelation_SFR_M}, a Spearman correlation test revealed a correlation coefficient of $\rho$ = -0.4 with a statistical significance of 8 $\sigma$, indicating  moderate anti-correlation between SFR$_{100}$ density and gas metallicity. Additionally, we computed the partial correlation coefficient at fixed stellar mass density to further investigate the observed trend. We found a significant negative partial correlation of $\rho$ = -0.40$^{+0.09}_{-0.08}$, confirming that at fixed stellar mass density, higher values of star formation rate density are associated with lower gas-phase metallicities. 

Our findings qualitatively align with the scenario in which a gas stream flows from one component of CR7 to the other (see Sec. \ref{Subsec3_kinematic}). In this case, the inflowing gas reduces the metal content of the ISM, thereby lowering the metallicity (see middle and right panels in Fig. \ref{fig:metalliticities}) and replenishing the gas reservoir of the accreting galaxy, thus increasing the SFR (see Fig. \ref{fig:sfr_10_100_Ha}). This scenario is consistent with significant amount of gas being dragged from CR7-C to CR7-A, as shown by extended ionised gas emission among these two satellites (see Figs. \ref{fig:mom_mapso3}, \ref{fig:narrow_broad_detached}). As a consequence, we observe a drastic decline of the gas reservoir, and thus the SFR in CR7-C (see Fig. \ref{fig:sfr_10_100_Ha}), with a net increase of the total gas-phase metallicity (see Fig. \ref{fig:metalliticities}). On the other hand, the effect on CR7-A, which is the main attractor of the gas mass inflowing from CR7-C, is that of an overall lower metal content and higher SFR. CR7-B appears not to be influenced by such interacting phenomena, showing an intermediate metal content and SFR (Tab. \ref{t.prospector} and Fig. \ref{fig:sfr_10_100_Ha}).

Similarly to the feature we observe in CR7-A, \cite{Arribas2023} found a metallicity gradient in a merging system at z $\sim$ 6.9 and interpreted it as due to accretion of metal poor circumgalactic medium (CGM) onto the main galaxy \citep[see also][]{Dekel2009, Cresci2010, delpino2024, Sarkar2024, venturi2024}. Additionally, \cite{Marconcini2024} reported a north-south metallicity gradient and a corresponding anti-correlation between $\Sigma_{M_{\star}}$ - $\Sigma_{SFR}$ in a star-forming galaxy at z $\sim$ 9.1 which could potentially be explained by the same scenario we propose here. Finally, \cite{Tripodi2024}, studied a sample of galaxies using NIRSpec/MSA data and observed similar trends in compact galaxies, further supporting this interpretation. From a theoretical perspective, \cite{bahe2015} studied the quenching of star formation in galaxy clusters from the GIMIC suite \citep{Crain2009} of cosmological hydrodynamical simulations and stated that quenching occurs earlier for satellite galaxies orbiting massive galaxies, due to tidal forces inducing gas loss.

\section{Conclusions}\label{Sec5_summary_conclusion}
We studied the ionised gas kinematics and excitation properties through emission line fitting in the z = 6.60425 system CR7, using JWST/NIRSpec high- and low-resolution IFU data cubes. Moreover, we have also performed a SED analysis to the integrated continuum emission of the three main UV bright companions obtaining their stellar masses, star-formation histories, and gas metallicities. The main results are summarized as follows:
\begin{itemize}
    \item The emission line multi-Gaussian fit of the high-resolution G395H data cube revealed the peculiar ionised gas morphology and kinematics of the three main satellites. Moment maps in Fig. \ref{fig:mom_mapso3} show blue-shifted extended emission of ionised gas connecting CR7-C and CR7-A, with projected velocities between $\sim$ -200 km s$^{-1}$ and 0 km s$^{-1}$. From our gas kinematic analysis we confirm that CR7-A, CR7-B, and CR7-C  have relative velocities smaller than 200 km s$^{-1}$ within a projected distance $\leq$ 6 kpc, and therefore we confirm that they are likely to represent an ongoing merging system. Moreover, we also discovered the presence of three smaller satellites with distinct narrow-line emission, possibly associated to interactions within this merging system. In particular two galaxies, CR7-A and CR7-C, show clear evidence of gravitational interactions with tidal tails connecting their main structures 

    \item We computed spatially resolved maps of emission line ratios and used tailored diagnostic diagrams to specifically discriminate the ionisation source at high redshift employing the \OIIIL/\OIIIL[4363] vs \OIIIL[4363]/\Hgamma line ratios. We found tentative evidence of AGN excitation southwards of CR7-A, co-spatial with enhanced values of \OIIIL/\Hbeta, further supporting the AGN ionisation scenario. In contrast to previous works we found that CR7-B is entirely associated with SF excitation, with no evidence of AGN ionisation, based on diagnostic diagrams. On the other hand, from the gas kinematic and line ratios (see Fig. \ref{fig:narrow_broad_detached}, \ref{fig:O3_over_Hbeta}) emerged the possibility of a high-ionisation source, possibly powering an outflow. 

    \item From the high-resolution data we computed the electron temperature (T$_e$) and found an average temperature of 1.6 $\times$ 10$^{4}$K. We estimated the gas-phase metallicity both with the direct T$_e$-method and SED fitting of low-resolution data and inferred higher metal content co-spatial with the AGN-ionised region, likely indicating the presence of a cavity emptied by an AGN-driven wind and then enriched with metals.

    \item We performed a detailed SED fitting analysis of the low-resolution data and re-constructed the SFH of CR7. Our findings suggest a tight interaction between CR7-A and CR7-C that occurred between 600 and 150 Myrs earlier, and impacted the evolution of both components. This interpretation is also supported by signs of gas exchange between  CR7-C and CR7-A, as shown by extended emission between these satellites. 

    \item We traced the \Lyalpha emission line from the low-resolution data and compared to the \Halpha emission to compute spatially resolved maps of the ionising photons escape fraction. We found that ionising photons are more likely to escape in the region where we found evidence of an outflow, consistent with the scenario of the wind clearing the gas along its path and thus favouring the escape of LyC photons towards the IGM, finally contributing to the re-ionisation. 
\end{itemize}

\begin{acknowledgements}
CM, FDE, and RM acknowledge support by the Science and Technology Facilities Council (STFC), by the ERC Advanced Grant 695671 ``QUENCH'', and by the UKRI Frontier Research grant RISEandFALL. CM also acknowledge the support of the INAF Large Grant 2022 ``The metal circle: a new sharp view of the baryon cycle up to Cosmic Dawn with the latest generation IFU facilities'' and of the grant PRIN-MUR 2020ACSP5K\_002 financed by European Union – Next Generation EU.
RM is further supported by a research professorship from the Royal Society.
SA, BRdP, and MP acknowledge support from the research project PID2021-127718NB-I00 of the Spanish Ministry of Science and Innovation/State Agency of Research (MICIN/AEI).
H{\"U} acknowledges support through the ERC Starting Grant 101164796 ``APEX''.
GCJ, AJB acknowledges funding from the "FirstGalaxies" Advanced Grant from the European Research Council (ERC) under the European Union’s Horizon 2020 research and innovation programme (Grant agreement No. 789056).
SC, EP, and GV acknowledge support by European Union's HE ERC Starting Grant No. 101040227 - WINGS.
IL acknowledges support from grant PRIN-MUR 2020ACSP5K\_002 financed by European Union – Next Generation EU.
GC acknowledges the support of the INAF Large Grant 2022 ``The metal circle: a new sharp view of the baryon cycle up to Cosmic Dawn with the latest generation IFU facilities''.
PGP-G acknowledges support  from  Spanish  Ministerio  de  Ciencia e Innovaci\'on MCIN/AEI/10.13039/501100011033 through grant PGC2018-093499-B-I00.
\end{acknowledgements}
\section*{Data Availability}
The JWST/NIRSpec data used in this work has been obtained within the NIRSpec-IFU GTO programme (program ID 1217, Observation 6) and have been publicly available since May 2, 2024. The data presented in this work will be shared upon reasonable request to the corresponding author.

\bibliographystyle{aa}
\bibliography{example} 


\begin{appendix} 
\section{Line ratios}\label{subsub_line_ratios}

\begin{figure}
	\includegraphics[width=\linewidth]{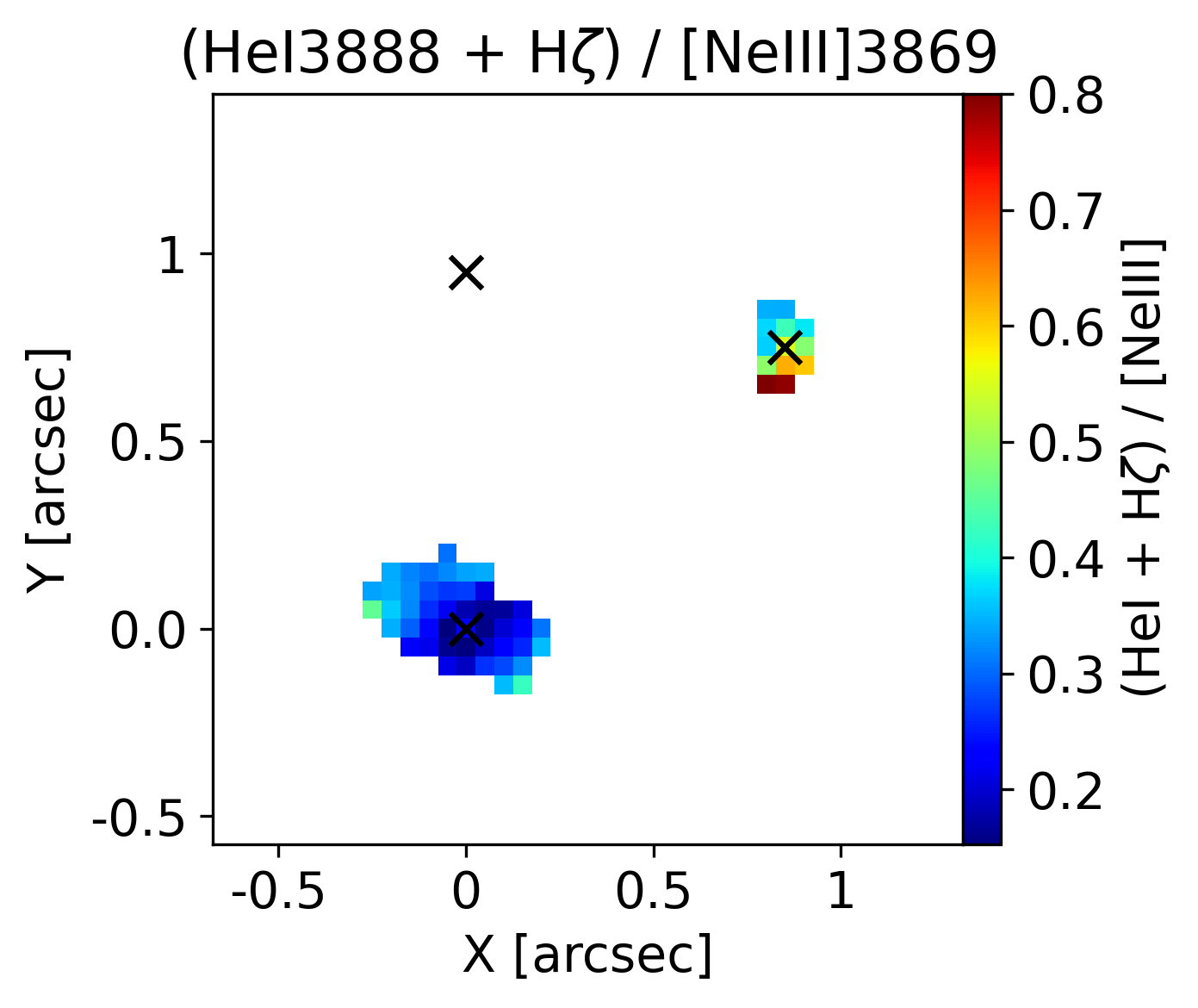}
    \caption{\HeIL[3888]+\Hzeta/\NeIIIL[3869] emission line ratio in CR7. Black solid lines are arbitrary \OIIIL levels. Spaxels with SN $\le$ 3 are masked. Emission lines are corrected for dust attenuation.}
    \label{fig:He1_Hz_Ne3}
\end{figure}

At $z>9.4$, \OIIIall is redshifted outside of the wavelength range covered by NIRSpec, preventing the use of these lines in various diagnostics of gas metallicity and ionization. In this context, \NeIIIL and \OIIall can be used as substitutes \citep[e.g.,][]{nagao2006,levesque+richardson2014,witstok+2021}.
In the context of \JWST observations, \citet{Tripodi2024} found that the \NeIIIL/\OIIall ratio in a sample of galaxies at $4<z<10$ decreases with galaxy radius, which could be due to inverse radial gradients in the gas metallicity. However, a complication of using \NeIIIL in the context of \JWST observations is that, with the spectral resolution of the NIRSpec prism, this line is blended with \HeIL[3888] and \Hzeta \citep{Tripodi2024}, which could also vary with radius, thus confounding the interpretation of the radial trends in \NeIIIL/\OIIall (which are actually radial trends in the ratio between the blend \NeIIIL+\HeIL[3888]+\Hzeta and \OIIall).
It is then useful to characterize the (\HeIL[3888]+\Hzeta)/\NeIIIL ratio in cases like CR7, where we have both the spectral and spatial information necessary to study spatial variations.
Fig. \ref{fig:He1_Hz_Ne3} shows the (\HeIL[3888] + \Hzeta)/\NeIIIL[3869] line ratio in CR7. It can be seen that the ratio increases outward. If this increase is common at $4<z<10$, then the results of \citet{Tripodi2024} would make the radial gradient in deblended \NeIIIL/\OIIall even more negative.
While a single galaxy as CR7-A is clearly not sufficient to draw general conclusion, this shows the potential of in-depth studies of well resolved galaxies to address questions raised by larger samples obtained with 1-d NIRSpec spectroscopy.

\end{appendix}
\end{document}